\documentclass[twocolumn,showpacs,preprintnumbers,amsmath,amssymb]{revtex4}
\usepackage{rotating}
\usepackage{makecell}
\usepackage{array,multirow}

\usepackage[normalem]{ulem}
\usepackage{graphicx,epsfig}
\usepackage{dcolumn}
\usepackage{bm}
\usepackage{xcolor}

\def\cD{{\cal D}}
\def\cN{{\cal N}}

\def\cR{{\cal R}}
\def\cB{{\cal B}}

\newcommand{\req}[1]{Eq.~(\ref{#1})}
\newcommand{\avg}[1]{\langle #1\rangle}

\newcommand{\fig}[1]{Fig.~\ref{#1}}

\DeclareMathOperator*{\argmin}{{\rm{argmin}}}

\makeatletter
\newcommand{\thickhline}{%
    \noalign {\ifnum 0=`}\fi \hrule height 1pt
    \futurelet \reserved@a \@xhline
}
\newcolumntype{"}{@{\hskip\tabcolsep\vrule width 2pt\hskip\tabcolsep}}
\makeatother

\newcommand{\cut}[1]{{}}
\usepackage{soul}
\newcommand{\snij}{\sigma^\nu_{ij}}
\newcommand{\snji}{\sigma^\nu_{ji}}
\newcommand{\snijs}{\sigma^{\nu *}_{ij}}

\newcommand{\tsnij}{\tilde{\sigma}^\nu_{ij}}

\newcommand{\tsmji}{\tilde{\sigma}^\mu_{ji}}
\newcommand{\tsmil}{\tilde{\sigma}^\mu_{il}}
\newcommand{\Iijs}{I^*_{ij}}
\newcommand{\cH}{{\cal H}}
\newcommand{\cC}{{\cal C}}
\newcommand{\fs}{f_{\rm s}}
\newcommand{\vs}{{\boldsymbol\sigma}}
\newcommand{\vsss}{{\boldsymbol\sigma}^{*}}
\newcommand{\vsssg}{{\boldsymbol\sigma}^{*}{(\gamma)}}

\newcommand{\vsssgr}{{\boldsymbol\sigma}^{*}{(\gamma_r)}}
\newcommand{\tss}{{\tilde{\boldsymbol \sigma}}^*}
\newcommand{\tns}{{\tilde{\boldsymbol \sigma}}^{\nu *}}

\newcommand{\tn}{{\tilde{\boldsymbol \sigma}}^{\nu}}
\newcommand{\tm}{{\tilde{\boldsymbol \sigma}}^{\mu}}
\newcommand{\vns}{{\boldsymbol \sigma}^{\nu *}}
\newcommand{\vli}{{\bf \Lambda}_i}
\newcommand{\lione}{\Lambda_i^{\mu}}
\newcommand{\linone}{\Lambda_i^{\setminus\mu}}
\newcommand{\vri}{{\bf R}_i}
\newcommand{\rione}{{R}_i^{\mu}}
\newcommand{\rinone}{{R}_i^{\backslash\mu}}
\newcommand{\sjione}{\sigma_{ji}^{\mu}}
\newcommand{\Ijinone}{I_{ji}^{\backslash\mu}}
\newcommand{\silone}{\sigma_{il}^{\mu}}
\newcommand{\Iilnone}{I_{il}^{\backslash\mu}}
\newcommand{\silones}{\sigma_{il}^{\mu*}}
\newcommand{\Iilnones}{I_{il}^{\backslash\mu*}}
\newcommand{\sjiones}{\sigma_{ji}^{\mu*}}

\newcommand{\Ijinones}{I_{ji}^{\backslash\mu*}}

\newcommand{\tE}{\tilde{E}}
\newcommand{\sone}{\sigma^{\mu}}
\newcommand{\Inone}{I^{\backslash\mu}}
\newcommand{\sones}{\sigma^{\mu*}}
\newcommand{\tsones}{\tilde{\sigma}^{\mu*}}
\newcommand{\Inones}{I^{\backslash\mu*}}
\newcommand{\Is}{I^*}
\newcommand{\tIs}{\tilde{I}^*}
\newcommand{\tmss}{\tilde{m}_s^*}
\newcommand{\tsone}{\tilde{\sigma}^\mu}
\newcommand{\D}{\Delta\cH}
\newcommand{\Dself}{\Delta\cH_{\rm selfish}}
\newcommand{\Dobed}{\Delta\cH_{\rm compliant}}

\begin{document}


\title{The Futility of Being Selfish  - The Impact of Selfish Routing on Uncoordinated and Optimized Transportation Networks}
\author{Ho Fai Po$^1$, Chi Ho Yeung$^{1}$\footnote{chyeung@eduhk.hk} and David Saad$^2$}
\affiliation{$^1$Department of Science and Environmental Studies, The Education University of Hong Kong, 10 Lo Ping Road, Taipo, Hong Kong. \\ $^2$The Nonlinearity and Complexity Research Group, Aston University, Birmingham B4 7ET, United Kingdom}

\date{\today}

\begin{abstract}
Optimizing traffic flow is essential for easing congestion. However, even when globally-optimal, coordinated and individualized routes are provided, users may choose alternative routes which offer lower \emph{individual} costs. By analyzing the impact of selfish route-choices on performance using the cavity method, we find that a small ratio of selfish route-choices improves the global performance of uncoordinated transportation networks, but degrades the efficiency of optimized systems. Remarkably, \emph{compliant users always gain in the former and selfish users may gain in the latter, under some parameter conditions}. The theoretical results are in good agreement with large-scale simulations. Iterative route-switching by a small fraction of selfish users leads to Nash equilibria close to the globally optimal routing solution. Our theoretical framework also generalizes the use of the cavity method, originally developed for the study of equilibrium states, to analyze iterative game-theoretical problems. These results shed light on the feasibility of easing congestion by route coordination when not all vehicles follow the coordinated routes.
\end{abstract}


\maketitle


\section{Introduction}
Traffic congestion is a major problem facing both metropolitan and highway road networks, which incurs high environmental and economic costs. While the economic costs on their own are mind-boggling (\$166Bn for the US in 2018~\cite{tamu2019}), the environmental costs through CO$_2$ emissions on climate change, and of fine particulate matter (PM$_{2.5}$), nitrogen dioxide (NO$_{x}$) and ozone, O$_{3}$ on public health are as significant (these three air pollutants alone are responsible for 400,000 premature deaths per year in the EU~\cite{particlesPublicHealth2018}). However, easing congestion through additional investment in infrastructure incurs its own momentary and environmental costs and one of the alternatives to mitigating congestion is traffic optimization through coordination~\cite{mogridge97}, supported by advances in information technology~\cite{ge20175g} and potentially the penetration of autonomous vehicles~\cite{urmson2008self}. While optimization algorithms for coordinating individual transportation routes to achieve a global objective~\cite{Noh02, Dobrin01, Bayati08, yeung12} show the global benefits, at the expense of a small increase in average path length of individual users~\cite{yeung2013physics, yeung2019coordinating}, it is clear that affecting road-user behavior is the key to their success~\cite{Shiftan2011,PRATO200965}.

Even when optimally coordinated individual routes are recommended, some road-users may choose alternative routes that incur lower individual costs~\emph{for them}. Dynamical selfish routing has been studied using game theory and operations research methodologies~\cite{fischer04,anshelevich09}, revealing the economic incentives which can suppress selfish behaviors~\cite{cole06} as well as the Nash equilibria in capacitated networks~\cite{correa04}. While most of these studies focused on the dynamics of \emph{individual route decisions}, the impact of selfish routing decisions on previously optimized transportation network has not yet been explored. Such analysis is of particular importance as it helps evaluate the potential benefit brought by global route coordination in future intelligent transportation system, via advanced information technology or self-driving vehicles.

Here, we introduce a transportation network model where users are given globally optimized individual routes from their start point to a common end point (e.g., city center or metropolitan hub), but some choose not to follow the recommended path but minimize their own cost. We apply the cavity approach developed for the studies of spin glasses~\cite{mezard87} to compute the re-routing probability by the selfish users and reveal their impact on the system globally. Both analysis and simulations demonstrate the benefit, \emph{for the individual}, of following the optimized suggestions. When selfish users are allowed to switch routes multiple times, the obtained solutions approach Nash equilibria. Finally, we simulate the model on the England highway network, where behaviors similar to those observed in random regular graphs and captured by the analytical predictions are observed. Our results reveal the impact of individual route choices on the global cost and the benefit in having coordinated routing in transportation networks even when some road-users do not follow the recommended routes. We also demonstrate how the cavity method, which was devised for studying equilibrium states, can be generalized to study iterative game-theoretical problems.

The paper is organized as follows: we will explain the problem and the corresponding model in Sec.~\ref{sec_model} and outline the derivation of the solution in Sec.~\ref{sec_analytic}. In Sec.~\ref{sec_results} we show the results obtained in synthetic and realistic network scenarios and examine numerically the impact of selfish decisions on the performance of the transportation system as a whole, as well as of selfish and compliant users. Summary and conclusions are presented in Sec.~\ref{sec_Conclusion}.
 
\section{Model}
\label{sec_model}
Consider $M$ vehicles denoted by $\nu=1,\dots,M$ traveling on a transportation network of $N$ nodes denoted by $i=1,\dots,N$. The density of vehicles is denoted by $\alpha=M/N$. Each vehicle $\nu$ starts from a random origin node $O_\nu$, and travels to a common and randomly drawn destination node $\cD$. We denote the route of vehicle $\nu$ on the link between nodes $i$ and $j$ by a variable $\snij$ as follows,
\begin{eqnarray}
\snij =
\begin{cases}
1, &\mbox{$\nu$ travels from $i\!\to\! j$}
\\
-1, &\mbox{$\nu$ travels from  $j\!\to\!i$}
\\0, &\mbox{if $\nu$ does not travel between $i$ and $j$,}
\end{cases}
\end{eqnarray}
such that $\snji = -\snij$. The total traffic flow from node $i$ to $j$ is $|I_{ij}| = \sum_\nu |\snij|$. Since traffic congestion occurs when more vehicles share the same road, one aims to minimize path-overlaps to suppress congestion. We therefore introduce the \emph{social travel cost} 
\begin{eqnarray}
\label{eq_H}
\cH(\vs| \gamma)=\frac{1}{M}\sum_{(ij)}\Big(\sum_\nu|\snij|\Big)^\gamma=\frac{1}{M}\sum_{(ij)}|I_{ij}|^\gamma,
\end{eqnarray}
where $\gamma>1$ and the cost increases non-linearly with the traffic flow to discourage the sharing of a link by multiple vehicles (other nonlinear costs can be accommodated within the same framework), $(ij)$ denotes the un-ordered combination of $i,j$, and the bold symbol $\vs = \{\snij\}_{\nu, i, j}$ denotes the vector over the variable $\snij$. For $\gamma = 1$, there is no interaction between routes and users merely minimize the length of their own routes irrespective of traffic 
\begin{eqnarray}
\cH(\vs|\gamma = 1)=\frac{1}{M}\sum_\nu\Big(\sum_{(ij)}|\snij|\Big)~.
\end{eqnarray}
To minimize $\cH(\vs| \gamma)$ with $\gamma > 0$, vehicles traveling on the same link always head in the same direction~\cite{yeung12}, i.e. either $\snij\ge 0$ or $\snij \le 0$ for all $\nu$ on the link $i\to j$, and the directed traffic flow between node $i$ and $j$ is given by 
\begin{eqnarray}
I_{ij}=\sum_\nu \snij,
\end{eqnarray}
such that vehicles go from $i$ to $j$ if $I_{ij}>0$ and vice versa. The route configuration which minimizes $\cH(\vs| \gamma)$ in~\req{eq_H} is found using a message-passing algorithm~\cite{yeung12}.

Since our primary goal is to study the impact of selfish routing decisions on  previously optimized transportation network, here we study a scenario in which the social travel cost $\cH(\vs| \gamma)$ has already been minimized, and the corresponding optimized route for each vehicle $\nu$ is identified and recommended. We denote the optimized route for vehicle $\nu$ by $\snijs=1$ if it includes travel from node $i$ to $j$, and $\snijs=0$ otherwise, where $\vsss$ is,
\begin{eqnarray}
\label{eq_vsss}
\vsss = \argmin_{\vs} \cH(\vs|\gamma),
\end{eqnarray}
Hence, the total recommended traffic from node $i$ to $j$ is denoted by $\Iijs = \sum_\nu\snijs$. We then consider a fraction $\fs$ of the $M$ users to be selfish. A selfish user $\nu$ follows the route $\tn$ that minimizes its own cost, defined as the impact of traffic on their own route - the \emph{individual travel cost} 
\begin{eqnarray}
\label{eq_selfishH}
\cC_\nu(\tn|\vsss, \gamma) = \sum_{i,j}|\tsnij|\left(1+|\Iijs - \snijs|\right)^{\gamma-1}.
\end{eqnarray}
It implies that user $\nu$ is aware of the (recommended) traffic induced by other vehicles in the network $|\Iijs - \snijs|$ and use it to minimize its overlap with the traffic. The cost nonlinearity $\gamma\!-\!1$ originates from the need to equate the sum of individual costs and the social cost of~\req{eq_H} as explained below. For $\gamma =2$  the individual cost reflect the cumulative traffic flow experienced by an individual over the chosen route.

This definition~(\ref{eq_selfishH}) leads to an interesting relationship between the social $\cH$ in~\req{eq_H} and individual travel costs $\cC_\nu$. We note that when all vehicles follow the recommended path, i.e. $\tsnij = \snijs$ for all $\nu$ and $i,j$, the summand in~\req{eq_selfishH} is given by
\begin{eqnarray}
|\snijs|\left(1+|\Iijs\!-\!\snijs|\right)^{\gamma\!-\!1} =
\begin{cases}
|\Iijs|^{\gamma\!-\!1}, &\mbox{when $|\snijs|\!=\!1$,}
\\
0, &\mbox{when $|\snijs|\!=\!0$,}
\end{cases}
\end{eqnarray}
such that $|\snijs|\left(1+|\Iijs - \snijs|\right)^{\gamma-1}=|\snijs||\Iijs|^{\gamma-1}$ and the sum of all individual travel costs is given by 
\begin{align}
\label{eq_equiv}
\sum_\nu \cC_\nu(\vs^{\nu *}|\vsss, \gamma) 
\!=\! \sum_\nu \sum_{i,j}|\snijs||\Iijs|^{\gamma-1} \!=\! \sum_{i,j}|\Iijs|^\gamma,
\end{align}
which is the social travel cost $\cH$. This relation implies that \emph{if all users follow the recommended routes, the sum of their individual costs is the same as the optimal social cost. However, if some users do not follow the recommended routes, the social travel cost can only remain unchanged or increase, since the recommended path configuration already minimizes the social cost.}

To quantify the impact of selfish decisions on the social travel cost, we measure the fractional change in $\cH$ due to re-routing by selfish users
\begin{eqnarray}
\label{eq_delta}
\D(\gamma_r, \gamma) = \frac{\cH(\tss(\gamma_r)|\gamma)-\cH(\vsssgr|\gamma)}{\cH(\vsssgr|\gamma)}
\end{eqnarray}
where $\vsssgr$ is the configuration of path recommended to the users which minimizes $\cH(\vs|\gamma_r)$; however, the real travel cost is characterized by $\cH(\vs|\gamma)$ where $\gamma$ is not necessarily equal to $\gamma_r$. 
On the other hand, the variables $\tss(\gamma_r)$ denote the configuration of routes after selfish users have re-routed from the recommended $\vsssgr$ to optimize their individual cost $\cC(\tss|\vsss, \gamma)$. In other words, $\tss(\gamma_r)=\{\tns(\gamma_r) \}_{\nu=1,\dots,M}$ and $\tns(\gamma_r)$ for vehicle $\nu$ is given by
\begin{align}
&\tns(\gamma_r) 
\\
&=
\begin{cases}
\vns(\gamma_r), &\mbox{for compliant users,} \nonumber
\\
\argmin_{\tn} \cC_\nu(\tn|\vsssgr, \gamma), &\mbox{for selfish users.} \nonumber
\end{cases}
\end{align}

In this paper, we reveal the impact of selfish route decisions by examining the quantity $\D(\gamma_r, \gamma)$, as well as similar quantities defined for both selfish and compliant users separately. Although our theoretical derivations in Sec.~\ref{sec_analytic} can accommodate all $\gamma_r, \gamma\ge 1$ in $\D(\gamma_r, \gamma)$, we will focus our studies on two scenarios, namely 
\begin{enumerate}
\item
the case with $(\gamma_r, \gamma)=(1, 2)$, i.e. when all users are originally recommended to follow their shortest path $\vsss(1)$, i.e. $\gamma_r=1$, but the social and individual costs are characterized by $\cH(\vs|2)$ and $\cC(\tss|\vsss(1), 2)$ respectively, both with $\gamma=2$ to discourage link-sharing by multiple vehicles;
\item
the case with $(\gamma_r, \gamma)=(2, 2)$, i.e. the original configuration of recommended path already minimizes the social cost $\cH(\vs|2)$, but selfish users re-route to a path $\tn\neq\vsss(2)$ to minimize their own cost $\cC(\tn|\vsss(2), 2)$; the re-routing could potentially increase the social cost as suggested by the relation~\req{eq_equiv}.
\end{enumerate}

\section{Analytical Solution}
\label{sec_analytic}

We employ tools used in the study of spin-glass systems to compute the switching probability of selfish users on a link, which is used for the computation of $\D(\gamma_r, \gamma)$ in~\req{eq_delta}. In Sec.~\ref{sec_optimal}, we first describe the approach developed in~\cite{yeung12} which gives the analytical solution of the system minimizing the social travel cost $\cH$. In Sec.~\ref{sec_selfish}, we follow a similar framework to derive a new analytic approach to describe the re-routing behavior of selfish users.

\subsection{Optimization of the social travel cost $\cH$}
\label{sec_optimal}

We first describe the framework developed in \cite{yeung12} to map a route optimization problem into a problem of resource allocation. In this case, we assign a transportation load $\Lambda_i$ to each node $i$, such that
\begin{eqnarray}
\label{eq_lambda}
\Lambda_i =
\begin{cases}
1, &\mbox{if $i = O_\nu, \exists\nu$;}
\\
-\infty, &\mbox{if $i = \cD$;}
\\
0, &\mbox{otherwise.}
\end{cases}
\end{eqnarray}
To ensure a path to the common destination is identified for each vehicle, each of them transfers the (unit) load from their origin to a common destination, which serves as a sink. In this case, we restrict all traffic flows to take up integer values, and the net resources $R_i$ on each node $i$ (except the destination) to be conserved as follows
\begin{eqnarray}
R_i = \Lambda_i + \sum_{j\in \cN_i} I_{ji} = 0
\end{eqnarray}
where $\cN_i$ is the set of neighboring nodes of $i$.

We then employ the cavity approach~\cite{mezard87} and assume that for the sparsely-connected networks studied only large loops exist, such that neighbors of node $i$ become statistically independent if it is being removed. At zero temperature, we express the optimized energy $E_{i\to l}(I_{il})$ of a tree network terminated at node $i$ as a function of the traffic flow $I_{il}$ from $i$ to $l$. Next, we write down a recursion relation to relate $E_{i\to l}(I_{il})$ to the energy $E_{j\to i}(I_{ji})$ of its neighbors $j$ other than $l$~\cite{yeung12}, given by
\begin{eqnarray}
\label{eq_prl_recur}
E_{i\to l}(I_{il}) = \min_{\left\{ \left\{I_{ji}\right\}\left| R_i=0\right\}\right.}
\Bigg[ |I_{il}|^\gamma\!+\!\sum_{j\in\cN_i\backslash l}E_{j\to i}(I_{ji}) \Bigg].
\end{eqnarray}
We further note that $E_{i\to l}(I_{il})$ is an extensive quantity of which the value of energy is dependent on the number of nodes in the network; we therefore define an intensive quantity $E^V_{i\to l}(I_{il})$ as
\begin{eqnarray}
\label{eq_prl_EV}
E^V_{i\to l}(I_{il}) = E_{i\to l}(I_{il}) - E_{i\to l}(0)
\end{eqnarray}
To obtain the analytical results of the system, we have to solve for the functional probability distribution $P[E^V(I)]$. By using the recursion relation in~\req{eq_prl_recur}, we can write a self-consistent equation in terms of $P[E^V(I)]$, by summing over all degrees $k$, in the form
\begin{align}
\label{eq_prl_prob}
P[E^V(I)]\!=&\!\int dk \frac{P(k) k}{\avg{k}} \int d\Lambda d\Lambda \prod_{j=1}^{k-1} \int dE_j^V P[E_j^V(I)] 
\nonumber\\
&\times \delta\Big(E^V(I)-\cR_{14}[\{E_j^V\}, \Lambda, I]\Big)
\end{align}
where $\cR_{14}$ denotes the right hand side of~\req{eq_prl_EV}, which indeed has to be computed by the recursion relation~\req{eq_prl_recur}; $P(k)$ and  $\avg{k}$ represent the degree distribution and its average, respectively. The analytical results obtained by numerically solving~\req{eq_prl_prob} can be found in \cite{yeung12}. In addition to providing analytical results for the routing system macroscopically,~\req{eq_prl_recur} can be used as an optimization algorithm to identify individualized optimal path configurations microscopically in real instances.

\subsection{Analysis of selfish re-routing behavior}
\label{sec_selfish}

To analyze the impact of re-routing by the selfish vehicles, we \emph{single out a vehicle} $\mu$ and analyze its re-routing behavior from the path configuration identified by~\req{eq_prl_recur}. In this case, we denote the resources on node $i$ by $\vli = (\lione, \linone)$, such that $\lione$ and $\linone$ denote the transportation load for vehicle $\mu$ and any other user respectively. Similar to~\req{eq_lambda}, $\vli$ given by
\begin{eqnarray}
\vli = 
\begin{cases}
(1,0), &\mbox{if $i = O_\mu$;}
\\
(0,1), &\mbox{if $i = O_\nu, \exists \nu\neq\mu$;}
\\
(-\infty, -\infty), &\mbox{if $i = \cD$;}
\\
(0,0), &\mbox{otherwise.}
\end{cases}
\end{eqnarray}
We further define the net resources on node $i$ to be $\vri = (\rione, \rinone)$, such that
\begin{align}
\rione &= \lione + \sum_{j \in {\cN}_i}\sjione,
\\
\rinone &= \linone + \sum_{j \in {\cN}_i}\Ijinone,
\end{align}
where $\rione$ and $\rinone$ denote the net resources for vehicle $\mu$ and for the other users, while $\sjione$ and $\Ijinone$ denote the flow of vehicle $\mu$ and the rest of the traffic. To ensure each vehicle has a path to the common destination, we restrict all flows to take up integer values and $(\rione, \rinone) = (0,0)$.

Here, we analyze a scenario in which the path configuration that optimizes the social travel cost $\cH(\vs|\gamma_r)$ is computed and known to all users; $\mu$ then re-routes to optimize its individual cost $\cC_\mu(\tm|\vsss, \gamma)$ based on the traffic in the recommended configuration. To achieve the goal, similar to Sec.~\ref{sec_optimal}, we define the optimized energy $E_{i\to l}(\silone, \Iilnone)$ of the tree network terminated at node $i$ to be a function of the traffic flow characterized by $\silone$ and $\Iilnone$ from $i$ to $l$. The rationale for introducing this energy is to separate the contribution of user $\mu$ on link $i\to l$ from the remaining traffic. We then write an equation relating $E_{i\to l}(\silone, \Iilnone)$ to $E_{j\to i}(\sjione, \Ijinone)$ from all neighboring node $j\in\cN_i$ of node $i$ excluding node $l$, given by
\begin{align}
\label{eq_recur1}
& E_{i\to l}(\silone,\Iilnone) = 
\\
& \min_{\{ \sjione,\Ijinone | \vri=(0,0)\}}
\!\left[ \left(|\silone|\!+\!|\Iilnone|\right)^{\gamma_r} \!+\!\sum_{j\in \cN_i \setminus l}\!E_{j\to i}(\sjione,\Ijinone)\right],
\nonumber
\end{align}
where the exponent $\gamma_r$ is adopted in the optimized path configuration identified and recommended to the users.

To characterize the re-routing by vehicle $\mu$, we introduce another energy $\tE_{i\to l}(\tsmil, \silones, \Iilnones)$, that considers the replacement of the recommended edge variable $\silones$ by the selfishly-optimized $\tsmil$, where $\Iilnones$ corresponds to the traffic resulting from the optimized paths for all other users, which minimize $\cH(\vs|\gamma_r)$. By using~\req{eq_recur1}, given a set of $(\silones, \Iilnones)$, one can express the corresponding optimal $(\sjiones, \Ijinones)$ from the neighboring node $j$ of node $i$ excluding node $l$ by
{\small
\begin{align}
\label{eq_sistar}
&\{\sjiones, \Ijinones\}_{j\in \cN_i\backslash l} =
\\
& \argmin_{\{ \sjione,\Ijinone | \vri=(0,0)\}}\!\left[ \left(|\silones|\!+\!|\Iilnones|\right)^{\gamma_r} \!+\!\sum_{j\in \cN_i \setminus l}\!E_{j\to i}(\sjione,\Ijinone)\right],
\nonumber
\end{align}
}
which is similar to~\req{eq_prl_recur} except for the separation of user $\mu$.
Thus, the values of $(\sjiones, \Ijinones)$ are indeed dependent on $(\silones, \Iilnones)$ and have already been computed in~\req{eq_recur1}. By using this relation, one can similarly write an equation to relate the energy $\tE_{i\to l}(\tsmil, \silones, \Iilnones)$, where the recommended path $\silones$ has been replaced by the selfishly-optimized $\tsmil$, to $\tE_{j\to i}(\tsmji, \sjiones, \Ijinones)$ for $j\in \cN_i \setminus l$, given by
{\small
\begin{align}
\label{eq_recur2}
&\tE_{i\to l}(\tsmil, \silones, \Iilnones) =  
\\
&\min_{\{ \{ \tsmji \} | \rione\!=\!0\}} 
\!\left[ |\tsmil| (1\!+\!|\Iilnones|)^{\gamma-1}\!+\!\hspace{-0.15cm}\sum_{j\in \cN_i \setminus l}\!\tE_{j\to i}(\tsmji, \sjiones, \Ijinones)\right].
\nonumber
\end{align}
}
We note that the exponent $\gamma$, instead $\gamma_r$, is used in computing the above individual cost. For clarity of presentation, we omit the dependence of $(\sjiones, \Ijinones)$ on $(\silones, \Iilnones)$ in~\req{eq_recur2}.

\begin{figure}
\centerline{
\epsfig{figure=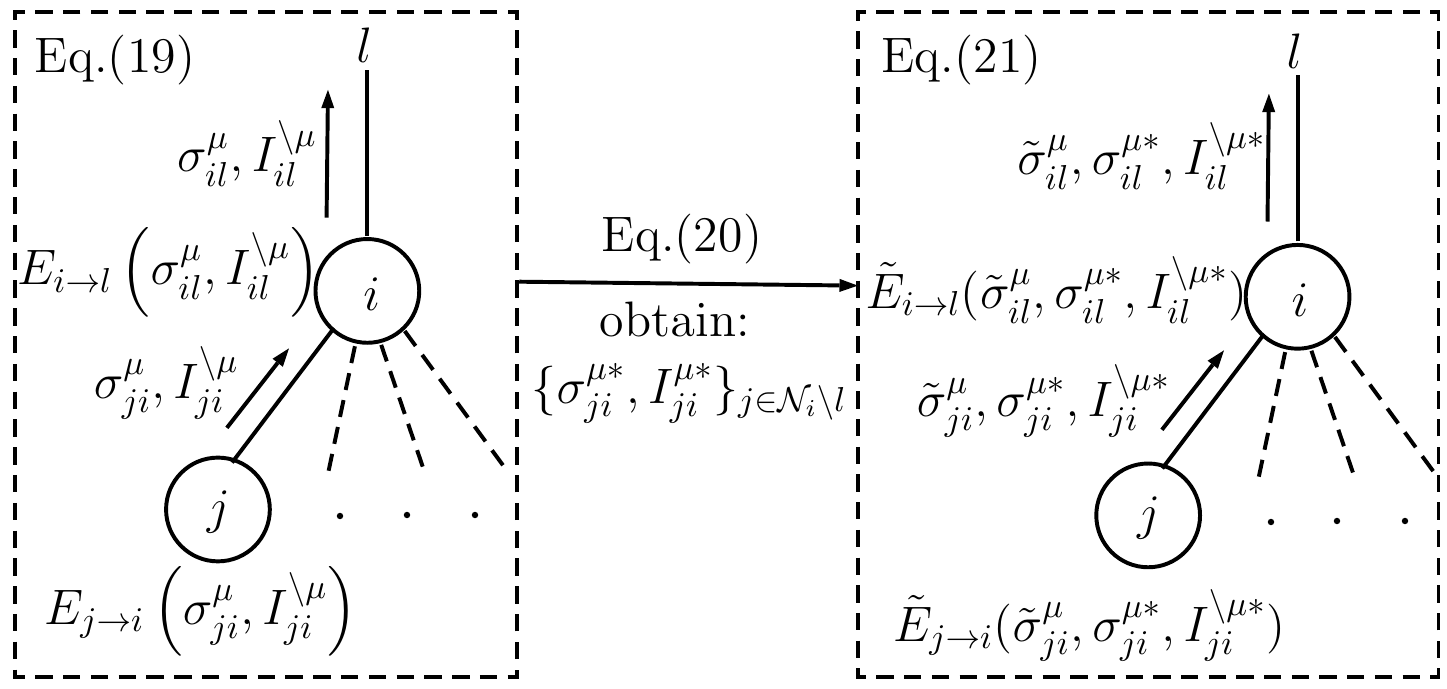, width=1\linewidth}
}
\caption{
A schematic diagram showing the recursion relation in Eqs.~(\ref{eq_recur1}) - (\ref{eq_recur2}), and the dependence of (\ref{eq_recur2}) on (\ref{eq_recur1}). The recursion relation is iterated simultaneously for all nodes to obtain the joint functional distribution $P[E^V, \tE^V]$.
}
\label{msgpassexplain}
\end{figure}

\emph{In summary:} The message passing process is described in \fig{msgpassexplain}. i) We use~\req{eq_recur1} to minimize the social travel cost $\cH(\vs|\gamma_r)$ for all users. ii) The corresponding optimal configuration of $(\sjiones, \Ijinones)$ from all neighbors $j$ for specific values $(\silones, \Iilnones)$ is given by~\req{eq_sistar}. iii) This information is used to compute $\tE$ using~\req{eq_recur2}, which is the \emph{optimized individual cost for vehicle} $\mu$.
Similar to~\req{eq_prl_EV}, we define the corresponding intensive quantities for $E$ and $\tE$ as
{\small
\begin{align}
\label{eq_EV}
&E^V_{i\to l}(\silone, \Iilnone)=E_{i\to l}(\silone, \Iilnone)\!-\!E_{i\to l}(0, 0), 
\\
\label{eq_tEV}
&\tE^V_{i\to l}(\tsmil, \silones, \Iilnones)\!
\nonumber\\
&\quad\quad\quad\quad\quad\quad=\tE_{i\to l}(\tsmil, \silones, \Iilnones)\!-\!\tE(0,0,0),
\end{align}
}
To compute $E^V$ and $\tE^V$ in the above equations, one has to iteratively follow the constrained optimization in Eqs.~(\ref{eq_recur1}) - (\ref{eq_recur2}).

To obtain the analytical solution of the re-routing behavior of vehicle $\mu$, we have to find the joint functional probability distribution $P[E^V(\sone, \Inone), \tE^V(\tsone, \sones, \Inones)]$; in the subsequent derivation, we denote this functional probability distribution as $P[E^V, \tE^V]$, omitting the arguments of $E^V$ and $\tE^V$ for clarity. In this case, we have to utilize both Eqs.~(\ref{eq_recur1}) and (\ref{eq_recur2}) to write down a self-consistent equation in terms of 
$P[E^V, \tE^V]$, given by
\begin{align}
\label{eq_prob}
&P[E^V, \tE^V] = \int dk \frac{P(k) k}{\avg{k}} \int d\Lambda d\Lambda 
\nonumber\\
&\quad\times \prod_{j=1}^{k-1} \int dE_j^V d\tE_j^V P[E^V_j, \tE_j^V]
\nonumber\\
&\quad\times \delta\Big(E^V(\sone, \Inone)-\cR_{22}[\{E_j^V\}, \Lambda, \sone, \Inone]\Big)
\nonumber\\
&\quad\times \delta\Big(\tE^V(\tsone, \sones, \Inones)-\cR_{23}[\{\tE_j^V\}, \Lambda, \tsone, \sones, \Inones]\Big),
\end{align}
where $\cR_{22}$ and $\cR_{23}$ correspond to the right hand side of the recursion relations in Eqs.~(\ref{eq_EV}) and (\ref{eq_tEV}) respectively. Equation (\ref{eq_prob}) can be solved numerically for $P[E^V, \tE^V]$.


\subsubsection{The probability of re-routing}

To derive various physical quantities of interest, we first compute the probability 
$p(\tsones, \sones, \Is)$, 
of vehicle $\mu$ re-routing from the original recommended traffic $\sones$ on a specific link to $\tsones$. For instance, $p(0,1,3)$ stands for the probability of vehicle $\mu$ switching its path away from a link with total traffic equal 3, which was part of the original optimized and recommended route. With the obtained joint probability $P[E^V, \tE^V]$, the expression for $p(\tsones, \sones, \Is)$ is given by
\begin{align}
\label{eq_Preroute}
&p(\tsones, \sones, \Is) = 
\nonumber\\
&\int dE_1^V d\tE_1^V P[E^V_1, \tE_1^V] \int dE_2^V d\tE_2^V P[E^V_2, \tE_2^V] \sum_{\Inones}
\nonumber\\
&\times\delta\Big((\sones, \Inones) 
\nonumber\\
&\quad\quad -\argmin_{(\sigma,I)}\big[E^V_1(\sigma,I) + E^V_2(-\sigma,-I) - |\sigma+I|^{\gamma_r}\big]\Big)
\nonumber\\
&\times\delta\Big(\Is - (\sones + \Inones)\Big)
\nonumber\\
&\times\delta\Big(\tsones - \argmin_{\tilde{\sigma}}\big[\tE^V_1(\tilde{\sigma},\!\sones,\!\Inones)
\nonumber\\
&\quad\quad+\tE^V_2(-\tilde{\sigma},\!-\sones,\!-\Inones) - |\tilde{\sigma}||1+\Inones|^{\gamma-1}\big]\Big)~,
\end{align}
which relies on evaluating the energy changes linked to the specified route switching.

\subsubsection{The cost of re-routing one selfish user}

By marginalizing over the joint probability $p(\tsones, \sones, \Is)$, we obtain $P(\Is)$, the probability that the original recommended traffic on a link is $\Is$, 
\begin{align}
\label{eq_PIs}
P(\Is) = \sum_{\tsones, \sones}p(\tsones, \sones, \Is)
\end{align}
We can then compute the optimal social travel cost $\cH(\vsssgr|\gamma)$ of the original system with recommended routes $\vsssgr$ by
\begin{align}
\label{eq_Hrec}
\cH(\vsssgr|\gamma)=\sum_{\Is} P(\Is) |\Is|^\gamma,
\end{align}
where $\vsssgr$ and $\Is$ are computed with respect to $\gamma_r$ by~\req{eq_Preroute} based on the recursion relation of~\req{eq_recur1}, while the exponent $\gamma$ reflects the fact that one computes the social cost of~\eqref{eq_H}. Similarly, we can also compute the cost $\cH(\tss(\gamma_r)|\gamma)$ where only \emph{one selfish user} has re-routed from the suggested path configuration, given by
\begin{align}
\label{eq_Hreroute}
&\cH(\tss(\gamma_r)|\gamma)
\\
&\hspace{1cm}=\sum_{\tsones, \sones, \Is} p(\tsones, \sones, \Is) (|\Is-\sones|+|\tsones|)^\gamma
\nonumber
\end{align}
Based on expressions (\ref{eq_Hrec}) and (\ref{eq_Hreroute}), we can compute the fractional change in cost, i.e. $\D(\gamma_r, \gamma)$ given by~\req{eq_delta}, for the system where there is only one re-routed vehicle. 

\subsubsection{The cost of re-routing multiple selfish users}
\label{sec_MFnoDistance}
To examine scenarios with multiple selfish users, one has to consider all permutations of choosing $t$ selfish users out of the $n$ users per link with re-routing probability $p$, which adheres to the Binomial distribution
\begin{align}
\cB(n, p, t) = {n \choose t} p^t (1-p)^{n-t},
\end{align}
with ${n \choose t} = n!/t!(n-t)!$ being the Binomial coefficient. We then denote $P(\tIs | \Is)$ to be the probability of having a traffic $\tIs$ on a link after selfish re-routing, given its original traffic  $\Is$. To facilitate the derivation, we define $\tmss$ to be the number of selfish users on a link after all selfish users in the system have re-routed their original recommended routes, and instead of $P(\tIs | \Is)$, we first write an expression for the probability $P(\tIs, \tmss | \Is)$ that depends on the number of rerouting decisions, given by
\begin{align}
\label{eq_PIms}
&P(\tIs, \tmss | \Is) = 
\hspace{-0.5cm}\sum_{m_s=\max(0, M\fs-(M-\Is))}^{\min(M\fs,\Is)} \hspace{-0.2cm}
\frac{{M\fs\choose m_s} {M(1-\fs) \choose \Is-m_s}}{{M\choose \Is}}
\nonumber\\
& \times\sum_{r=0}^{m_s}\cB\left(m_s, r, \frac{p(0,1,\Is)}{p(0,1,\Is) + p(1,1,\Is)}\right)
\nonumber\\
& \times\sum_{s=0}^{M\fs-m_s}\cB\left(M\fs-m_s, s, \frac{p(1,0,\Is)}{p(1,0,\Is)+p(0,0,\Is)}\right)
\nonumber\\
& \times\delta_{\tIs,\Is+(s-r)}\delta_{\tmss, m_s -r +s}.
\end{align}
All combinations of route changes are weighed with the appropriate combinatorial factor, with respect to the number of route-changes given the number of selfish users, and the probability of the Bernoulli distribution reflects the probability of changing route choices on a link.

With $P(\tIs, \tmss | \Is)$ computed, the probability $P(\tIs | \Is)$ is given by marginalizing over $\tmss$
\begin{align}
\label{eq_PI}
P(\tIs | \Is) = \sum_{\tmss = 0}^{M\fs}P(\tIs, \tmss | \Is)~.
\end{align}

Having obtained $P(\tIs | \Is)$, 
we can compute the global social travel cost $\cH(\tss(\gamma_r)|\gamma)$ after re-routing, given by 
\begin{align}
\cH(\tss(\gamma_r)|\gamma) = \sum_{\tIs} |\tIs|^\gamma \sum_{\Is} P(\tIs | \Is) P(\Is).
\end{align}
Using \req{eq_PIms} we can compute our main quantity of interest, i.e. the fractional change $\D(\gamma_r, \gamma)$ in the social cost given by~\req{eq_delta}. To the impact specifically on selfish and compliant users we compute the \emph{travel cost averaged over the selfish users}
\begin{align}
\label{eq_Hselfish}
&\cH_{\rm selfish}(\tss(\gamma_r)|\gamma) = \frac{1}{M\fs}\sum_{\tIs, \tmss} \tmss|\tIs|^{\gamma-1}
\nonumber\\
&\hspace{3cm}\times\sum_{\Is} P(\tIs, \tmss|\Is) P(\Is)~,
\end{align}
and the \emph{travel cost averaged over compliant users}
\begin{align}
\label{eq_Hobedient}
\cH_{\rm compliant}(\tss(\gamma_r)|\gamma) = &\frac{1}{M(1-\fs)}\sum_{\tIs, \tmss} (\tIs-\tmss)|\tIs|^{\gamma-1} 
\nonumber\\
&\times\sum_{\Is }P(\tIs, \tmss|\Is) P(\Is)~.
\end{align}
We note that the sum of the cost of the selfish and the compliant users is the total cost of the system after re-routing, i.e. $M\fs\cH_{\rm selfish} + M(1-\fs)\cH_{\rm compliant} = M\cH$. We can then define the fractional changes $\Dself(\gamma_r, \gamma)$ and $\Dobed(\gamma_r, \gamma)$ for selfish and compliant users respectively, similar to $\D(\gamma_r, \gamma)$ in~\req{eq_delta}.

\subsubsection{Effect of distance from the universal destination}
\label{sec_mf_distance}

We remark that the calculation in Sec.~\ref{sec_MFnoDistance} and the resulting probability $P(\tIs|\Is)$ in~\req{eq_PI} do not consider the heterogeneity of links in the network. For instance, a link with an original traffic $\Is=0$ would be more likely to have $\tIs>0$ after re-routing if it is closer to the common destination. To improve the equations of Sec.~\ref{sec_MFnoDistance} we incorporate the distance between the link of interest and the common destination within the derivation.

We define $d_i$ to be the minimum distance between node $i$ and the common destination; with $d_i=0$ implying that node $i$ is the common destination. We then modify Eqs.~(\ref{eq_recur1}) and (\ref{eq_recur2}), replacing the cavity energy functions $E_{i\to l}(\silone,\Iilnone)$ and $\tE_{i\to l}(\tsmil, \silone,\Iilnone)$ by the modified cavity functions $E_{i\to l}(\silone,\Iilnone|d_i)$ and $\tE_{i\to l}(\tsmil, \silone,\Iilnone|d_i)$, respectively, which incorporate a given distance to destination $d_i$. In the modified Eqs.~(\ref{eq_recur1}) and (\ref{eq_recur2}), we update the variable $d_i$ by 
\begin{align}
d_i = 1 + \min_{j\in \cN_i\backslash l }\{ d_j \}
\label{eq_d}
\end{align}
Next, we modify~\req{eq_Preroute} to compute the re-routing probability $p(\tsones, \sones, \Is|d)$ given the minimum distance $d$ to the common destination,
\begin{align}
d = \min(d_1, d_2),
\end{align}
where $d_1$ and $d_2$ correspond to the distance variables of $E_1, \tE_1$ and $E_2, \tE_2$ respectively in~\req{eq_Preroute}.

Using $p(\tsones, \sones, \Is|d)$, we modify~\req{eq_PIms} to obtain $P(\tIs, \tmss | \Is, d)$ and consequently $P(\tIs | \Is, d)$. The cost of the system after re-routing is thus given by
\begin{align}
\cH(\tss(\gamma_r)|\gamma) = \sum_{\tIs} |\tIs|^\gamma \sum_{\Is, d} P(\tIs| \Is, d) P(\Is, d),
\end{align}
where the joint probability $P(\Is, d)$ is obtained from the modified~\req{eq_Preroute}. Moreover, the travel cost averaged over the selfish users is given by
\begin{align}
\label{eq_Hselfish2}
\cH_{\rm selfish}(\tss(\gamma_r)|\gamma)=&\frac{1}{M\fs}
\sum_{\tIs, \tmss} \tmss|\tIs|^{\gamma-1} 
\nonumber\\
&\times\sum_{\Is, d}  P(\tIs,\!\tmss|\Is,\!d)P(\Is,\!d),
\end{align}
and the travel cost averaged over the compliant users is 
\begin{align}
\label{eq_Hobedient2}
\cH_{\rm compliant}(\tss(\gamma_r)|&\gamma) = \frac{1}{M(1\!-\!\fs)}\hspace{-0.1cm}\sum_{\tIs,\!\tmss} (\tIs\!-\!\tmss)|\tIs|^{\gamma-1}
\nonumber\\
& \times\sum_{\Is, d}  P(\tIs,\!\tmss|\Is,\!d)P(\Is,\!d) ~.
\end{align}
With both costs computed, we can examine whether various groups of users, i.e. users on average, selfish or compliant users, benefit after the selfish re-routing.

\section{Results} 
\label{sec_results}

As mentioned in Sec.~\ref{sec_model}, we mainly study two scenarios: (i) $(\gamma_r, \gamma)=(1,2)$ and (ii) $(\gamma_r, \gamma)=(2,2)$. The analytical results we obtained by the mean-field approach of Sec.~\ref{sec_mf_distance} which considers the distance from the common destination are compared with simulation results, and the results from a semi-analytical approach in which the re-routing probabilities $p(\tsones, \sones, \Is|d)$ are obtained from simulations instead of~\req{eq_Preroute}; the empirical probabilities are then inserted in the subsequent equations for the computation of the system's behavior. In general, the analytical and semi-analytical approaches exhibit very similar results, suggesting that the re-routing probabilities $p(\tsones, \sones, \Is|d)$ are estimated accurately in the analysis.

\begin{figure*}
\centerline{
\begin{tabular}{c}
\epsfig{figure=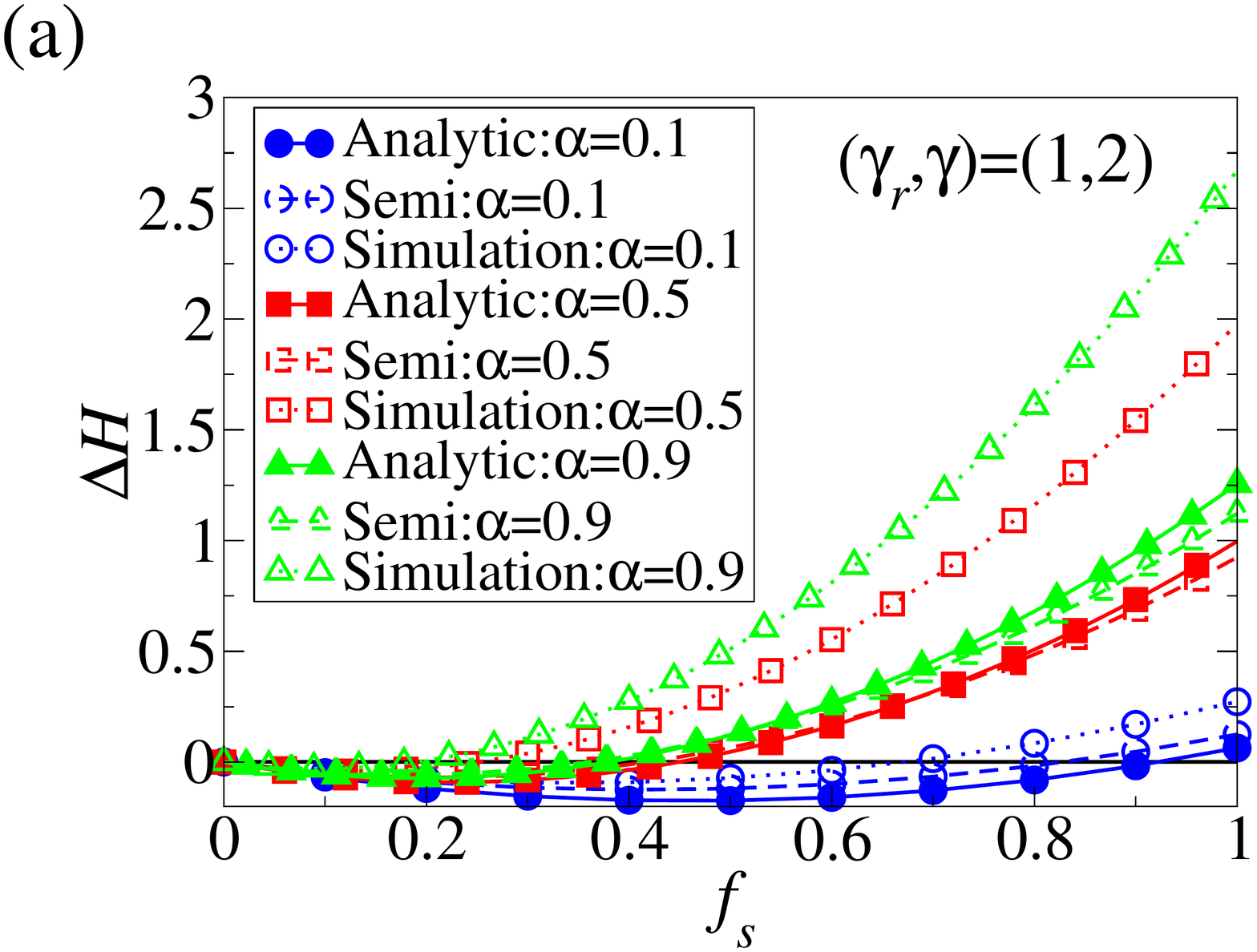, width=0.333\linewidth}
\epsfig{figure=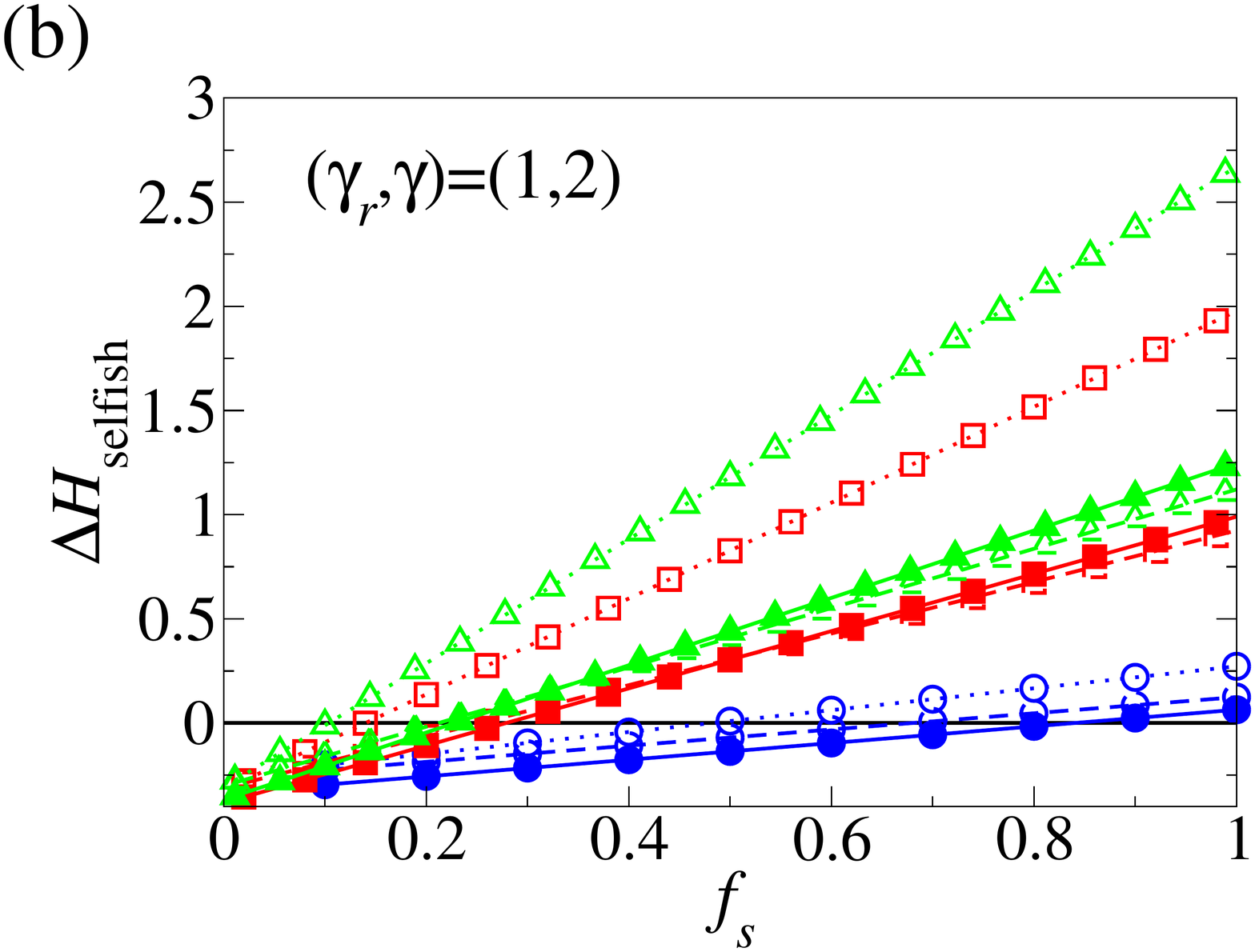, width=0.333\linewidth} 
\epsfig{figure=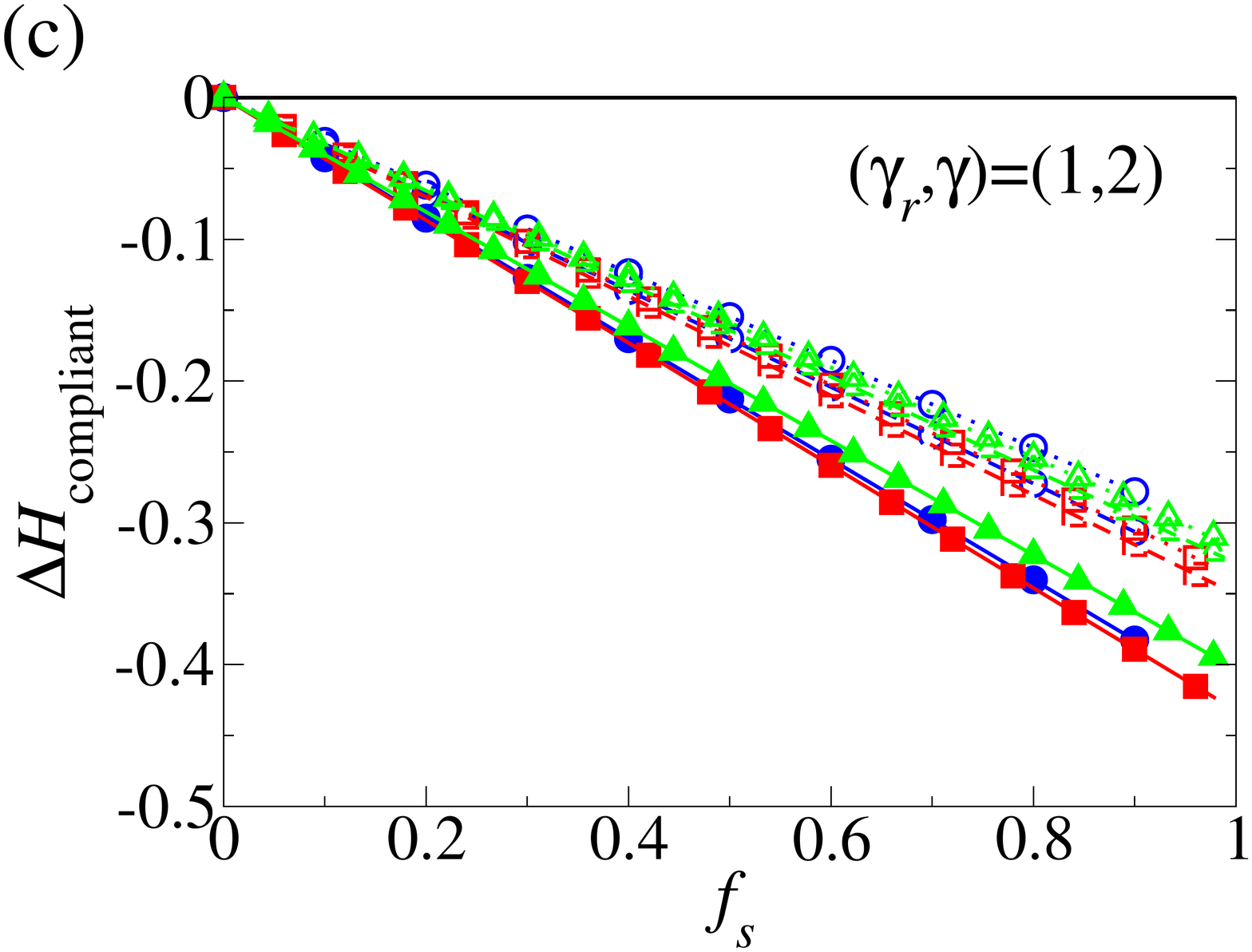, width=0.333\linewidth} \\
\epsfig{figure=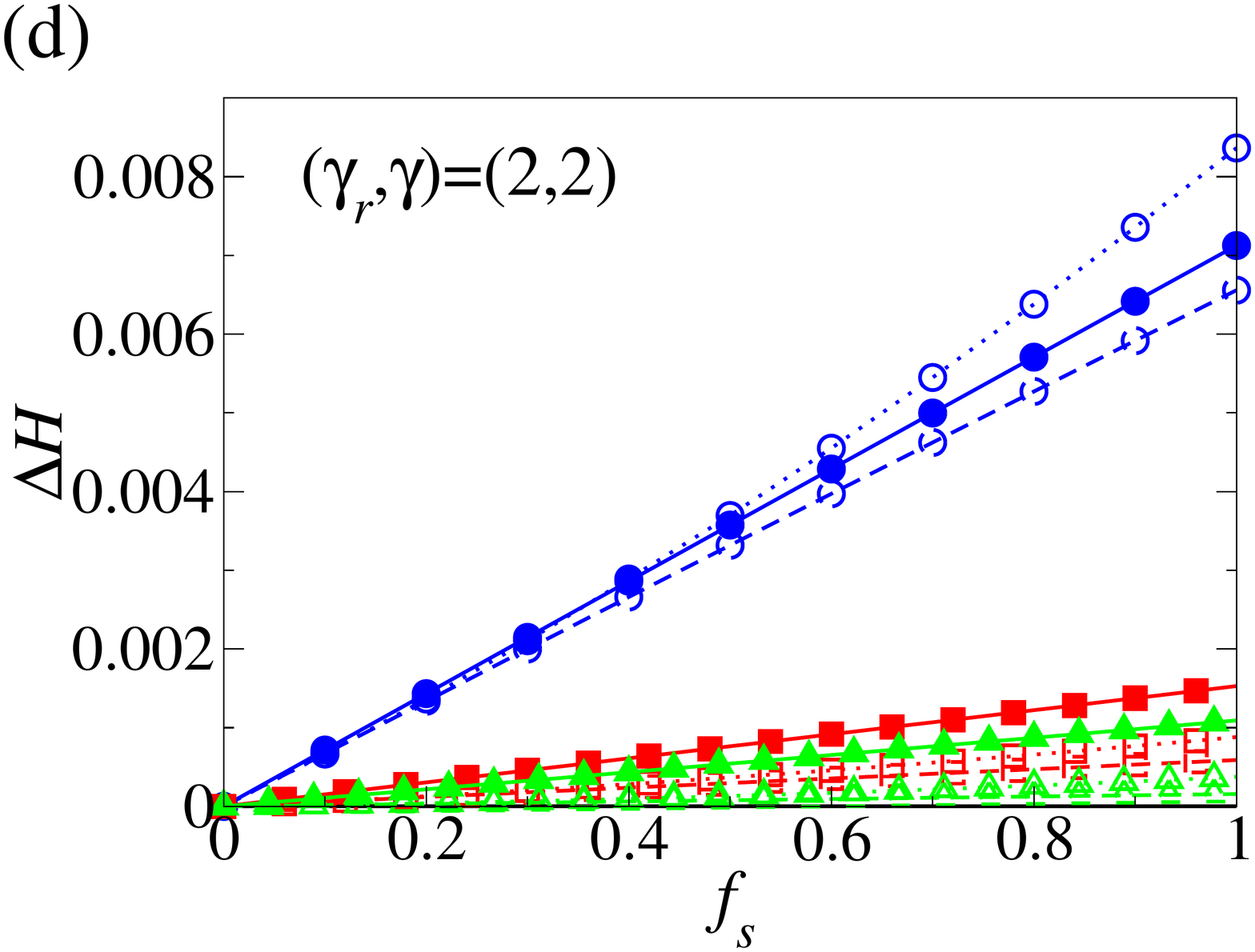, width=0.333\linewidth}
\epsfig{figure=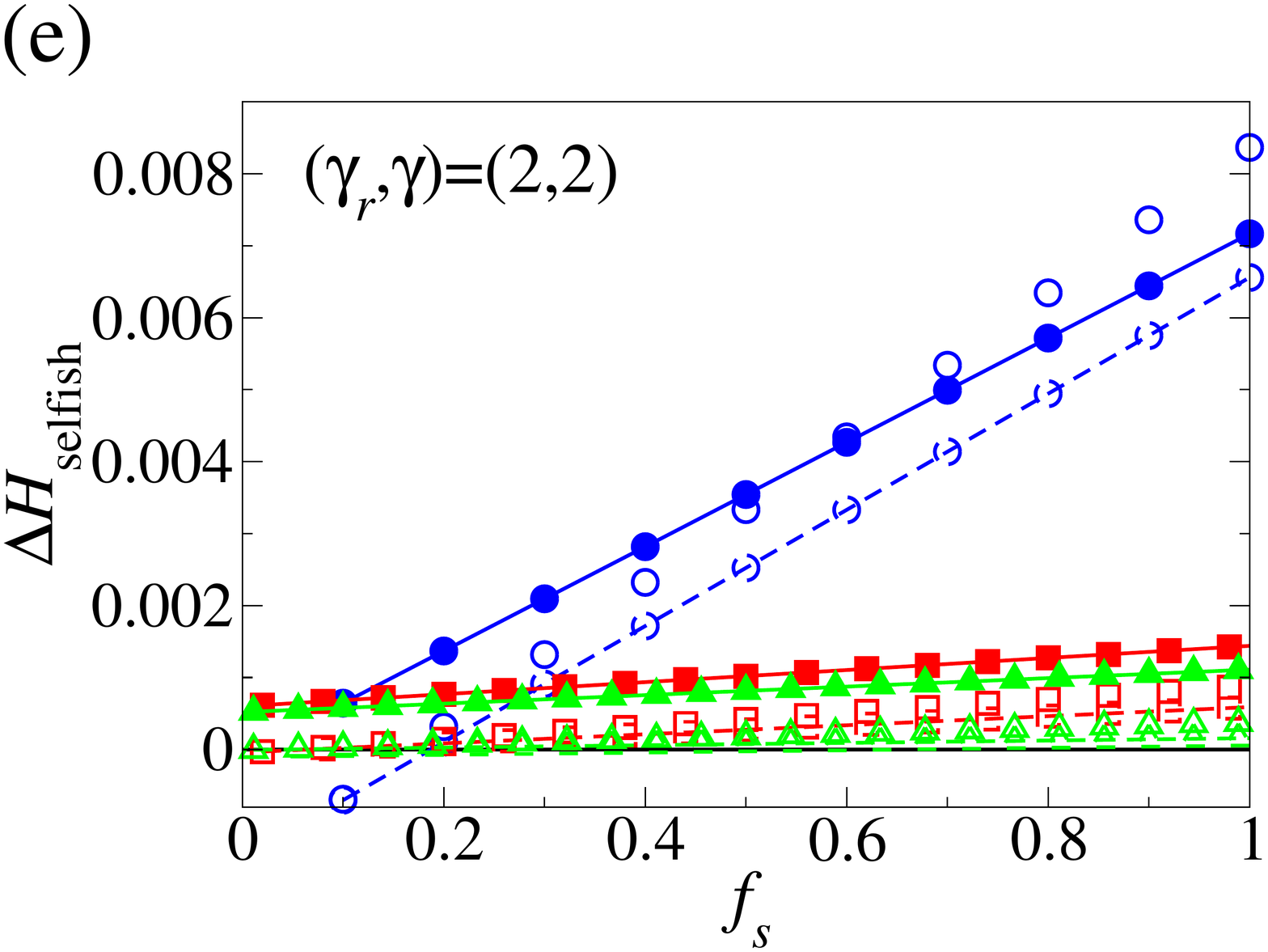, width=0.333\linewidth} 
\epsfig{figure=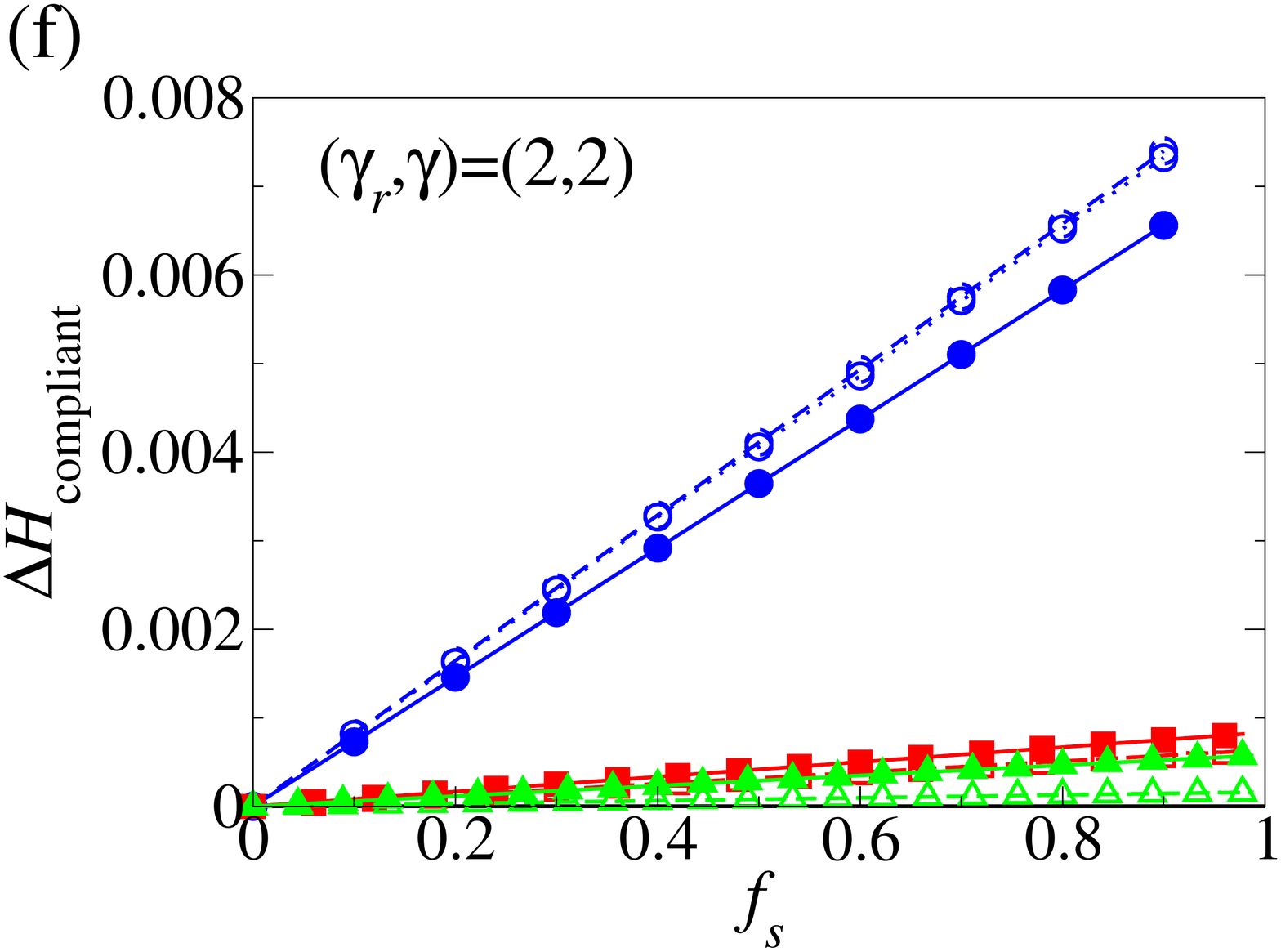, width=0.333\linewidth} \\
\end{tabular}
}
\caption{
The fractional change in travel costs (a) $\D$, (b) $\Dself$ and (c) $\Dobed$ as a function of $\fs$, averaged over all users,  selfish and  compliant users, respectively, for $(\gamma_r, \gamma)=(1,2)$. The corresponding results for cases with $(\gamma_r, \gamma)=(2,2)$ are shown in (d), (e) and (f). The simulation results are obtained on random regular graphs with $N=100$ and $k=3$ for vehicle density $\alpha=M/N=0.1, 0.5$ and $0.9$ averaged over 1000 realizations. The analytical and semi-analytical results are also shown. 
}
\label{FC}
\end{figure*}

\begin{figure}
\centerline{
\begin{tabular}{c}
\epsfig{figure=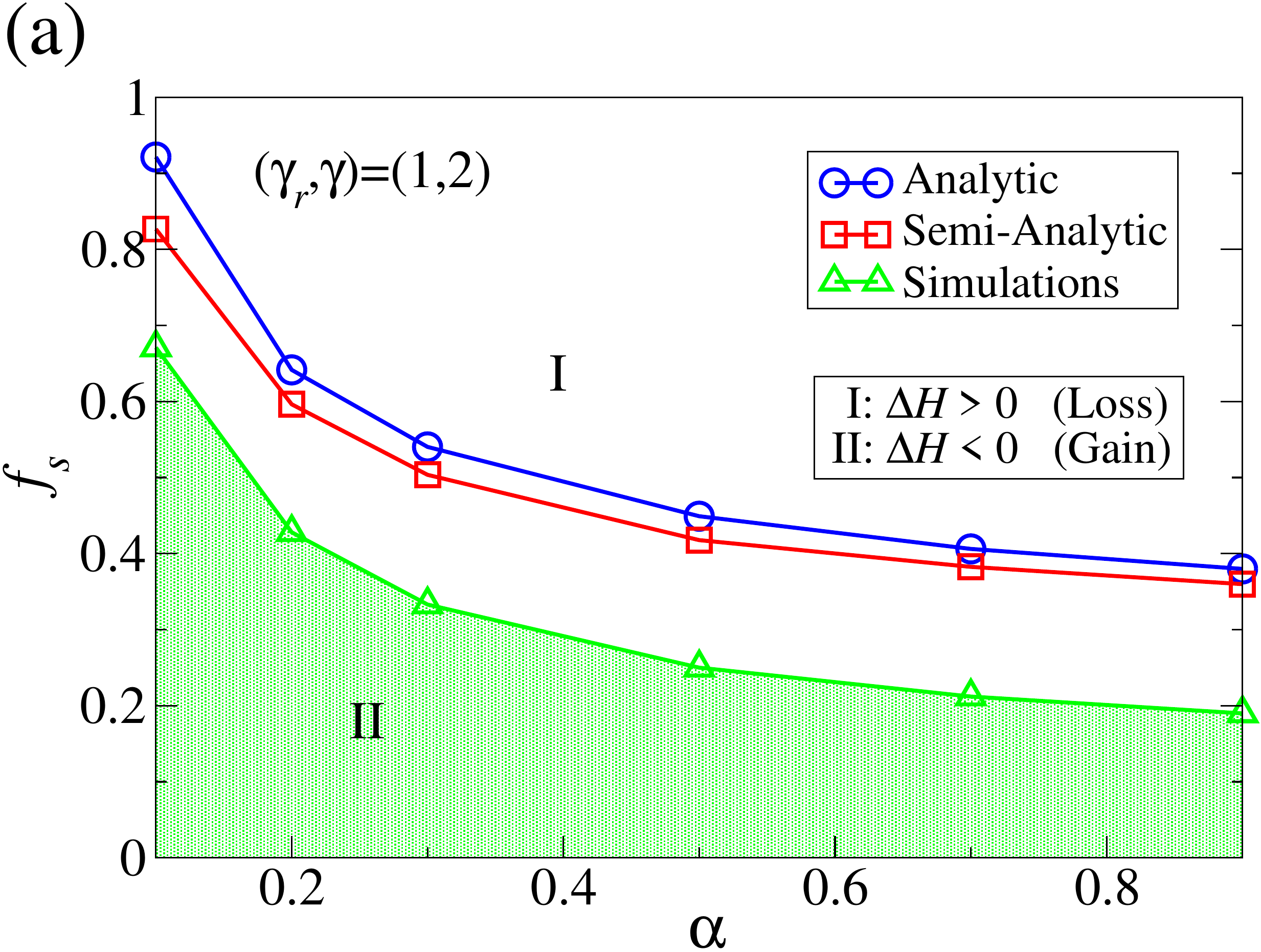, width=0.9\linewidth} \\
\epsfig{figure=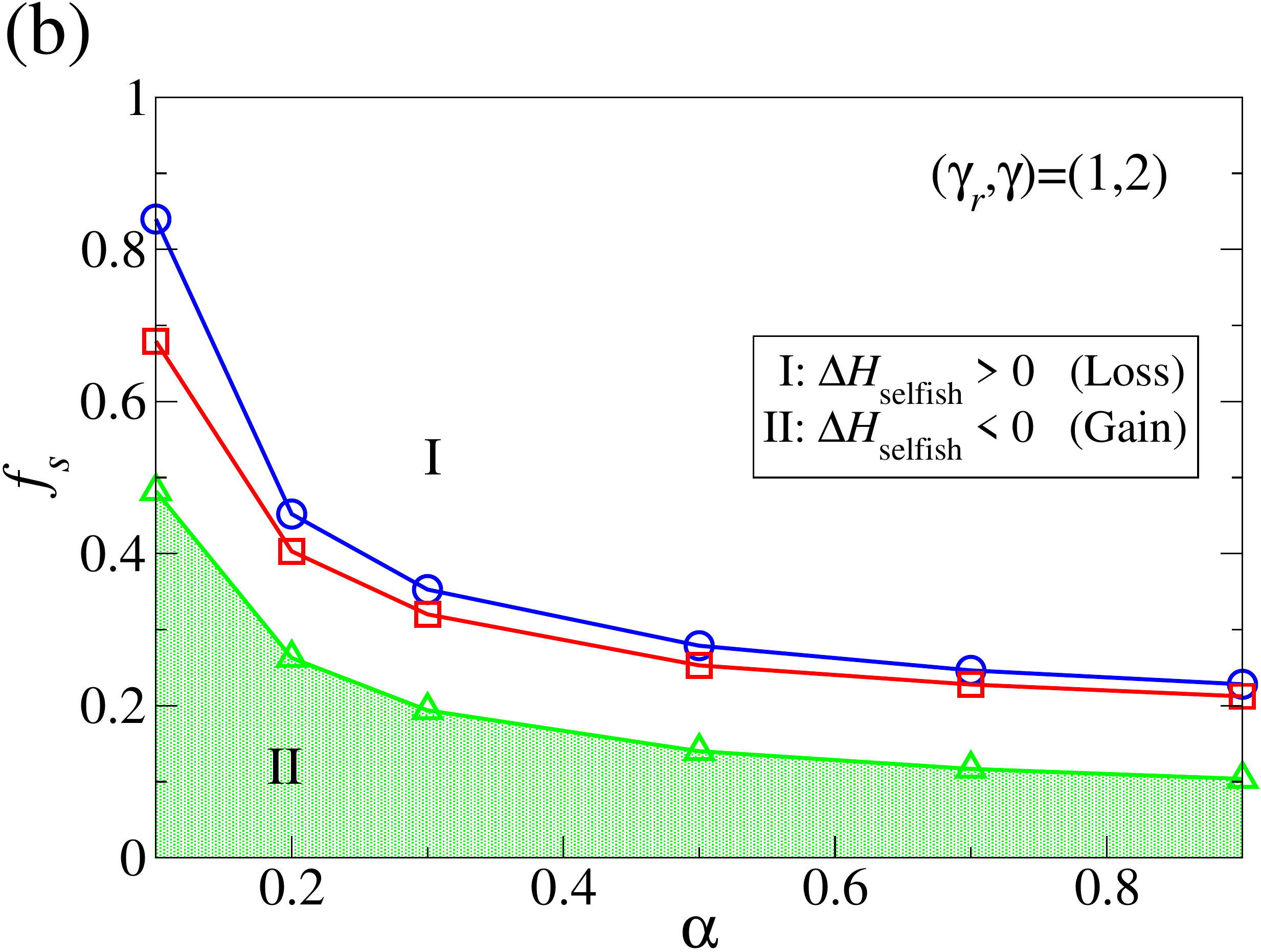, width=0.9\linewidth} \\
\epsfig{figure=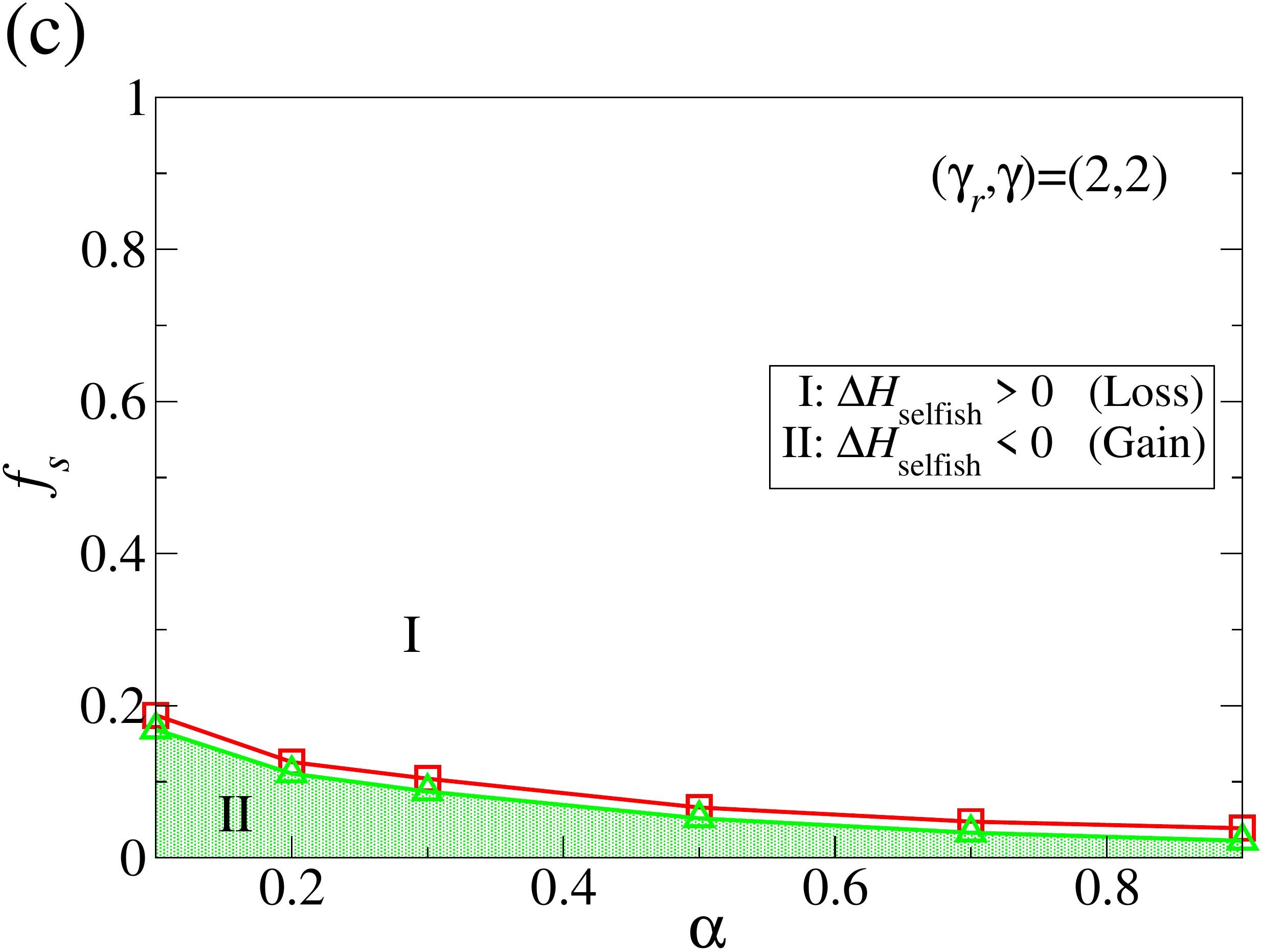, width=0.9\linewidth} \\
\end{tabular}
}
\caption{
The regimes with negative or positive (a) $\D(1,2)$, (b) $\Dself(1,2)$ and (c) $\Dself(2,2)$ are shown in terms of the fraction of selfish users $\fs$ and vehicle density $\alpha$. The predictions of the corresponding regime boundary by the analytical and semi-analytical approaches are shown. 
}
\label{phase}
\end{figure}

\subsection{Recommended shortest path - $(\gamma_r, \gamma)=(1,2)$}
\label{sec_ga12}
We first examine the case of $(\gamma_r, \gamma)=(1,2)$, in which all users are originally recommended to follow their shortest paths, but the social cost is quadratic in traffic flow and
consequently the individual cost incurs traffic flow cost linearly along its route. This scenario may correspond to an ordinary daily route choices without coordination, in which users just follow their shortest path but some may change their route after getting prior information about traffic loads.

As shown in \fig{FC}(a), for the cases of $\alpha = M/N=0.1, 0.5$ and $0.9$, the results from all the three approaches show that $\D(1,2)$, the fractional change in the social travel cost averaged over all users, first becomes negative and then increases back to a positive value as $\fs$ increases. This implies that when users originally follow their shortest path, a small fraction of selfish users is beneficial to the system as they occupy less-used roads freeing up heavily used roads, consequently $\D$ becomes negative. However, the system's cost increases when the fraction of selfish users increases, as they correlatively occupy the previously less-used roads, leading to congestion. Although the analytical results capture the negative and positive regime as well as the trends of the simulation results, there are discrepancies. We believe that the discrepancies come from the mean-field nature of~\req{eq_PIms}, since the analytical and semi-analytical results show a good agreement which suggests that the analytical approach estimates well the re-routing probabilities $p(\tsones, \sones, \Is|d)$ in simulations. The reason is arguably the high variability of costs and congestion levels that emerge from the randomness of both topology and travel start points, which the mean-field approach does not capture well. 

In addition, we see in \fig{FC}(a) that the lower the vehicle density $\alpha$, the larger the value of $\fs$ beyond which $\D$ becomes positive. To better reveal the benefits brought by the selfish re-routing, we show in \fig{phase}(a) the parameter regime in terms of $\alpha$ and $\fs$ in which the system gains (i.e. $\D<0$) after the selfish users re-routed. As we can see, as $\alpha$ increases, the regime with $\D<0$ decreases, implying that a suitable fraction of selfish users is beneficial to the system if all users originally just follow their shortest path. Similar to \fig{FC}(a), the analytical and the semi-analytical results capture the trend in the simulation results, but show discrepancies. These results imply that although the analytical approaches do not identify the exact regime boundary in the case of $(\gamma_r, \gamma)=(1,2)$, they predict the correct picture. 

Next, we examine the change in travel cost over the selfish and compliant user groups. As shown in \fig{FC}(b), we see that $\Dself$, the fractional change in cost for the selfish users, increases from a negative to a positive value when $\fs$ increases. The smaller the fraction of selfish users is the more negative the value of $\Dself$ is, suggesting that fewer selfish users gain more from the re-routing as they can better exploit less-used roads; this benefit vanishes when as the number of selfish users increases since their correlated re-routings generate congestion on the selected roads. As shown in \fig{phase}(b), the critical fraction of selfish users beyond which they start to lose decreases as the vehicle density $\alpha$ increases. More interestingly, \fig{FC}(c) shows that the \emph{compliant users always gain for every}  $\fs>0$, implying that the presence of selfish users is beneficial to  compliant users which follow their shortest paths.

\subsection{Quadratic loss optimization - $(\gamma_r, \gamma)=(2,2)$}
\label{sec_ga22}
In the case with $(\gamma_r, \gamma)=(2,2)$, users are recommended to follow the optimal configuration of paths which already minimizes the social cost, hence selfish re-routing is likely to increase it. As shown in \fig{FC}(d), the results from all the three approaches show that $\D > 0$ for all $\fs>0$ as expected, and the more the selfish users there are, the larger the cost. On the other hand, we also see that the results obtained from the analytic and the semi-analytic approaches show a good agreement with simulation results, better than that in the case of $(\gamma_r, \gamma)=(1,2)$ in \fig{FC}(a). This is arguably since in the optimized social cost case, traffic load is already balanced, in spite of the inherent variability induces by the randomness in topology and route starting points, and therefore the mean-field approach represents better the true localized and individualized fields, and probabilities.

Interestingly, although any changes must be detrimental to the system's cost, selfish users may gain in their individual cost as shown in \fig{FC}(e) at the expense of others. As we can see, when vehicle density $\alpha$ is small, $\Dself < 0$ when the fraction of selfish users is small, and $\Dself$ becomes positive when $\fs$ increases. This reflects the ability of a small fraction of users to exploit less-used routes selfishly, a strategy that backfires as their fraction increases, leading to correlations and congestion. 

We further show in \fig{phase}(c) the parameter regime within which selfish users gain, defined as the critical fraction of selfish users $\fs$ for a given vehicle density $\alpha$. This fraction $\fs$ decreases as $\alpha$ increases, implying that a smaller fraction of selfish users can gain when the vehicle density increases. The regime where $\Dself<0$ in the case of $(\gamma_r, \gamma)=(2,2)$ shown in \fig{phase}(c), is much smaller than that of the case $(\gamma_r, \gamma)=(1,2)$ shown in \fig{phase}(b), implying that less selfish users can gain in an initially-optimized system ($\gamma_r=2$) compared to the un-coordinated system ($\gamma_r=1$). This is in agreement with the previous results, since the variability of route loads in the un-coordinated routing allows for selfish users to secure improved re-routing. On the other hand, as shown in \fig{FC}(f), compliant users always lose due to the actions of selfish users, unlike the case of $\gamma_r=1$, but not as much in most $\fs$ and $\alpha$ values.

In summary, in cases where users are recommended to follow paths which already minimize the social cost, re-routing by the selfish users results in a higher social cost, increases the cost for compliant users, but selfish users themselves may gain \emph{if they are a small minority}.

\subsection{Multiple rounds of selfish re-routing and Nash equilibrium}
\label{sec_multipleRound}

\begin{figure}
\centerline{
\epsfig{figure=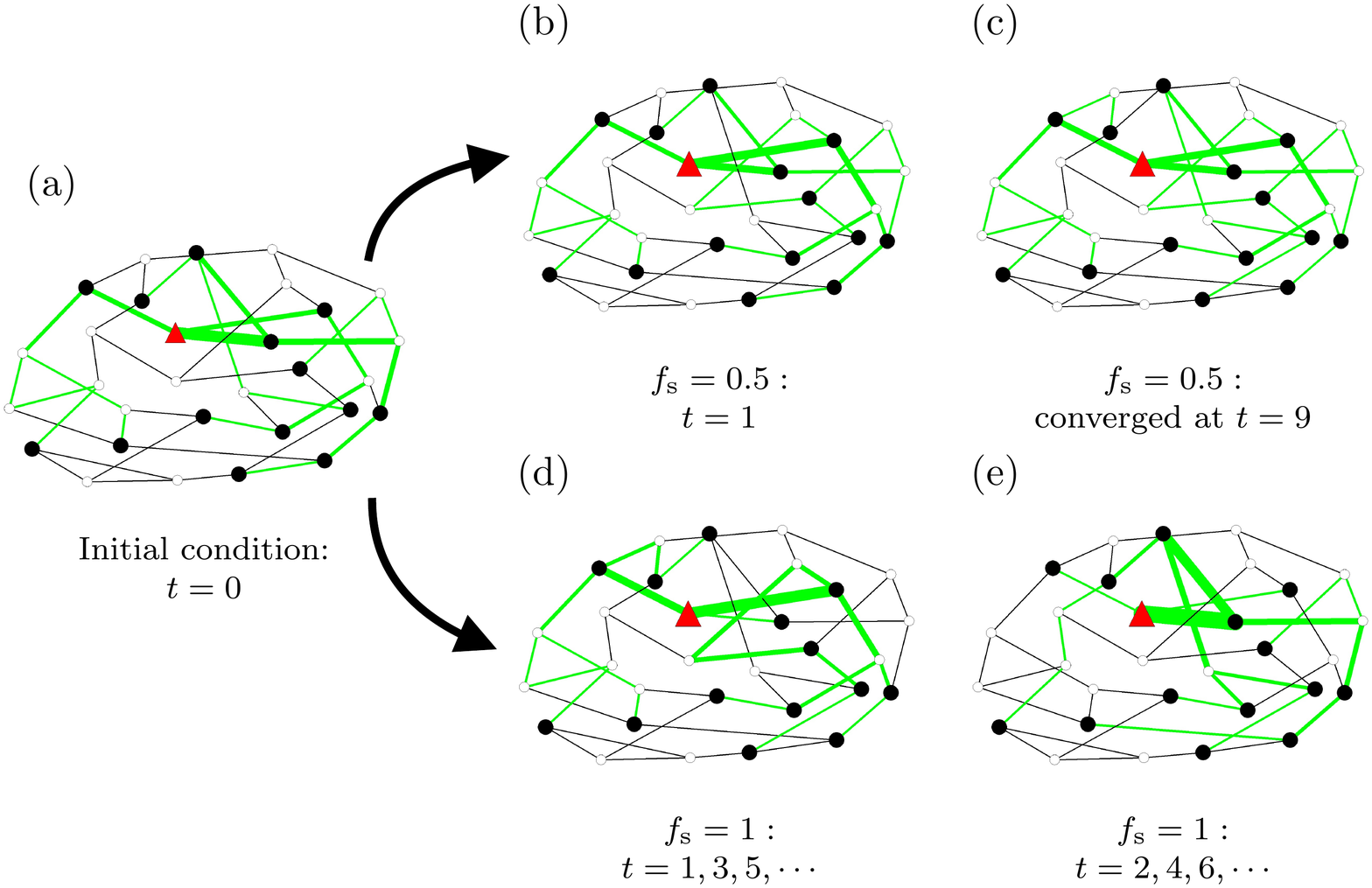, width=0.9\linewidth}
}
\caption{
An example of consecutive rounds of selfish re-routing by $M=14$ users on a network with $N=30$ nodes. The origins of users and their common destination are shown by the filled circles and triangle, respectively. Links with non-zero traffic are shown in green, with width proportional to the size of the flow. (a) All users travel via their shortest path to the destination at $t=0$. (b) Half of the users re-route at $t=1$ i.e. $\fs=0.5$. (c) The system arrives at a Nash equilibrium state at $t=9$. (d) All users re-route to minimize their own cost at $t=1$, i.e. $\fs=1$; as we can see, some links close to the destination have low or zero traffic flow. (e) At $t=2$, selfish users switch to what were less occupied links at $t=1$, leaving other links underloaded. The system then switches back and forth between the configurations in (d) and (e) repeatedly.
}
\label{nasheg}
\end{figure}

\begin{figure}
\centerline{
\begin{tabular}{cc}
 \multicolumn{1}{l}{(a)} &  \multicolumn{1}{l}{(b)}  \\
\epsfig{figure=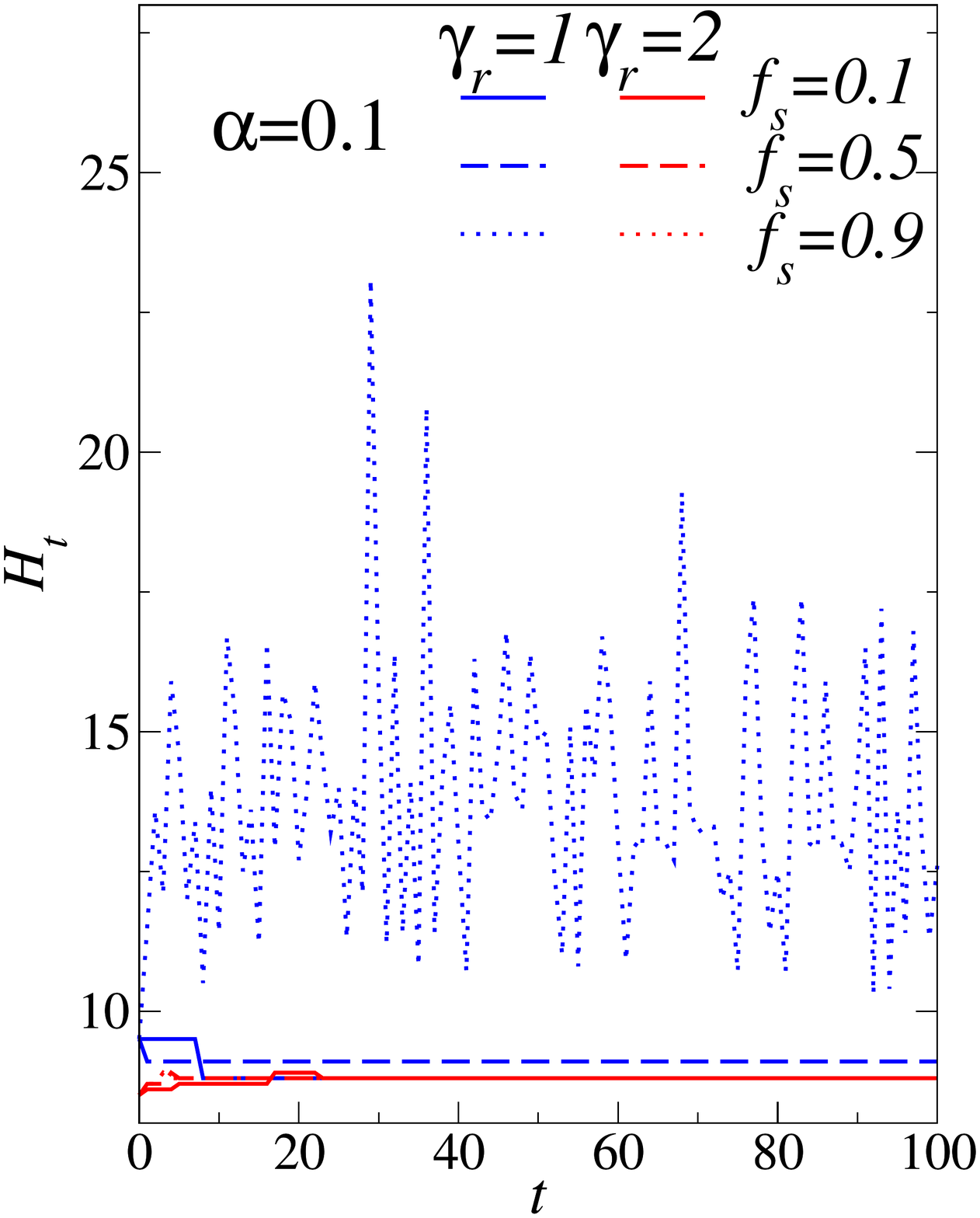, width=0.49\linewidth} &
\epsfig{figure=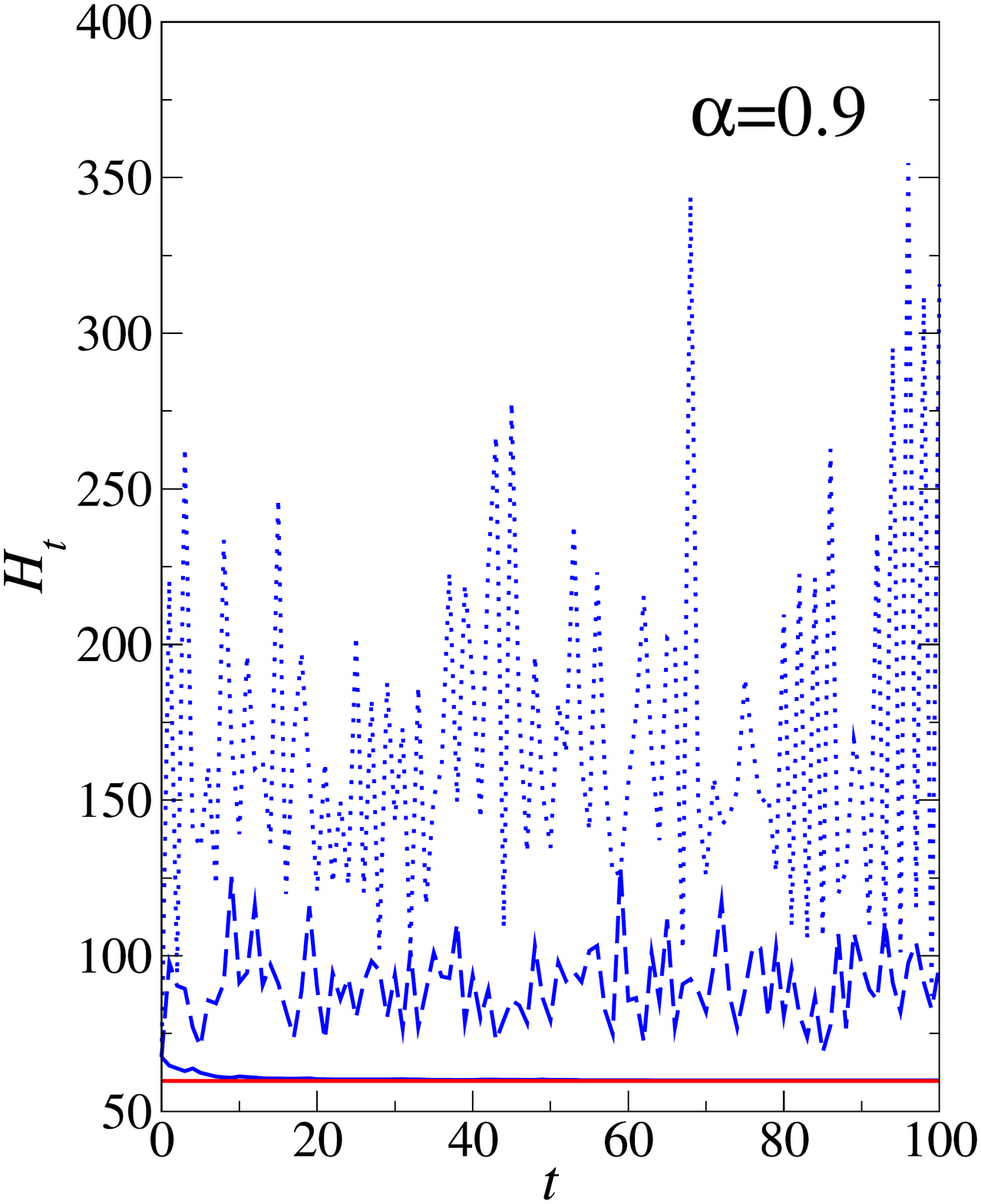, width=0.49\linewidth} \\
 \multicolumn{1}{l}{(c)} &  \multicolumn{1}{l}{(d)}  \\
\epsfig{figure=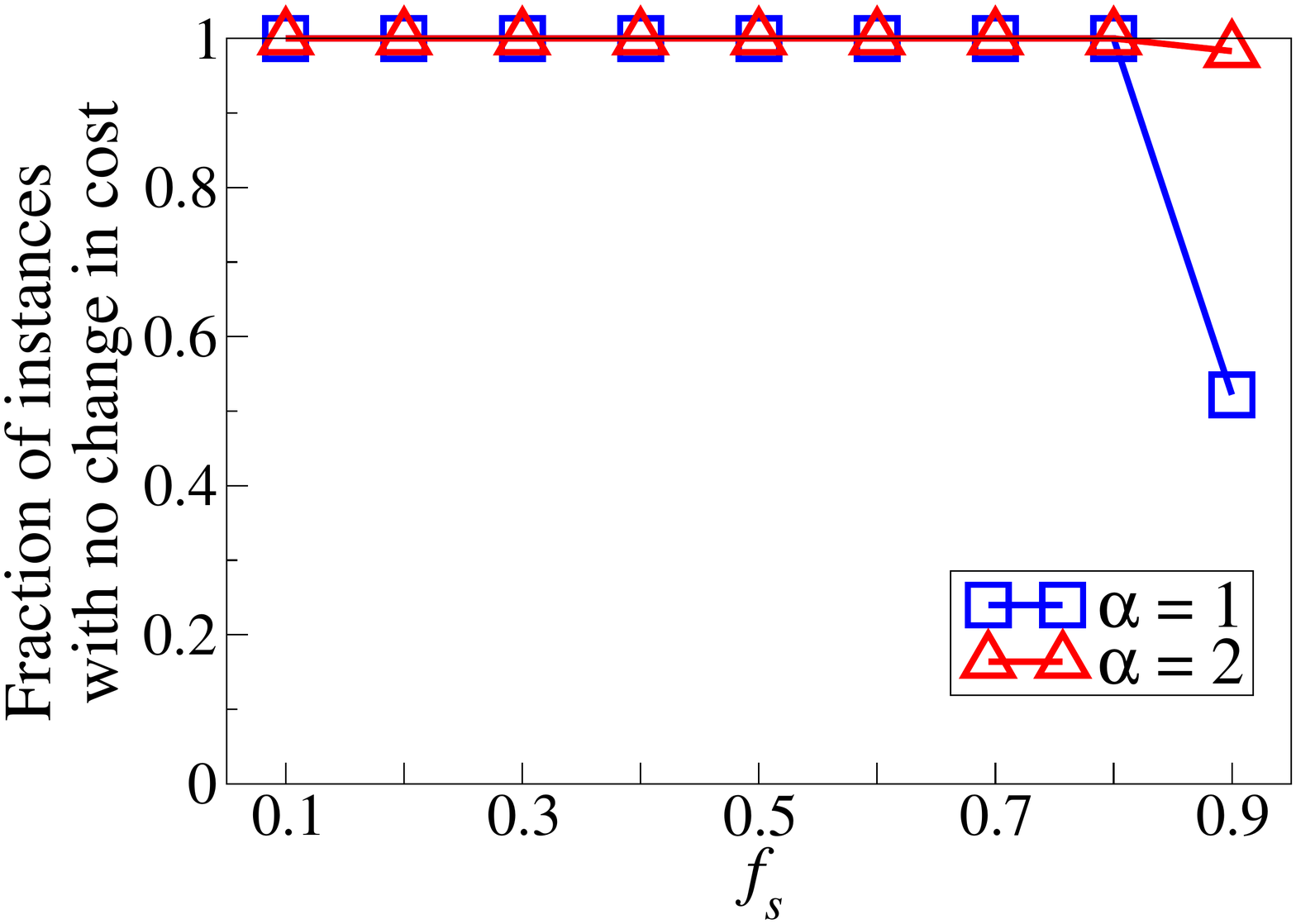, width=0.49\linewidth} &
\epsfig{figure=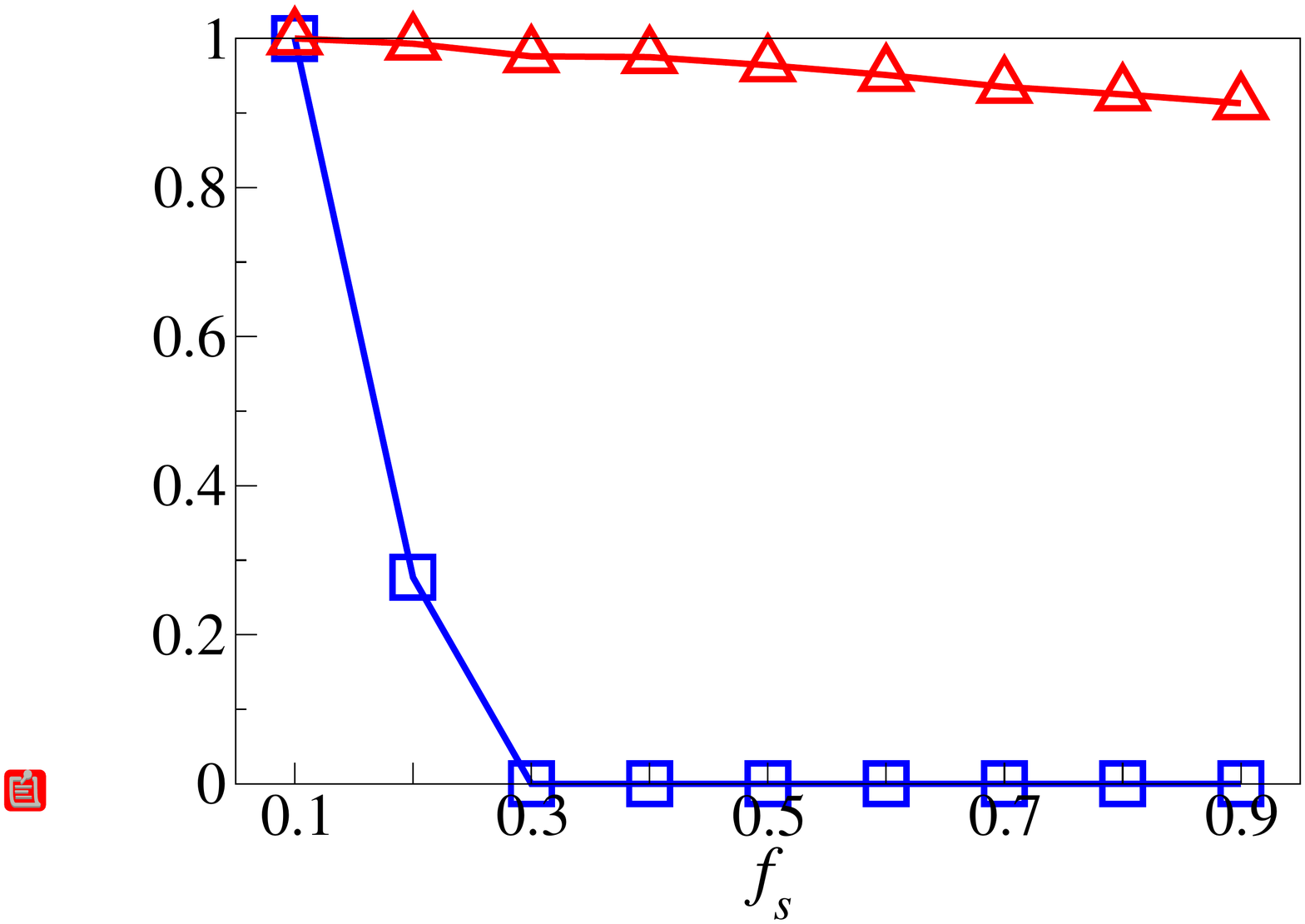, width=0.49\linewidth} \\
\end{tabular}
}
\caption{
(a, b) The social travel cost $\cH$ as a function of $t$, the number of selfish re-routing rounds, measured on specific instances of random regular graphs with $N=100$, $k=3$, $\gamma_r =1,2$ and various $\fs$ for (a) $\alpha=0.1$ and (b) $\alpha=0.9$.
(c, d) The fraction of instances which show no change in travel cost $\cH$ for the last consecutive 2500 steps in a simulation with 5000 steps, for (c) $\alpha=0.1$ and (d) $\alpha=0.9$.
}
\label{nashtimeseries}
\end{figure}

\begin{figure}
\centerline{
\begin{tabular}{cc}
\multicolumn{1}{l}{(a)} &  \multicolumn{1}{l}{(b)} \\
\epsfig{figure=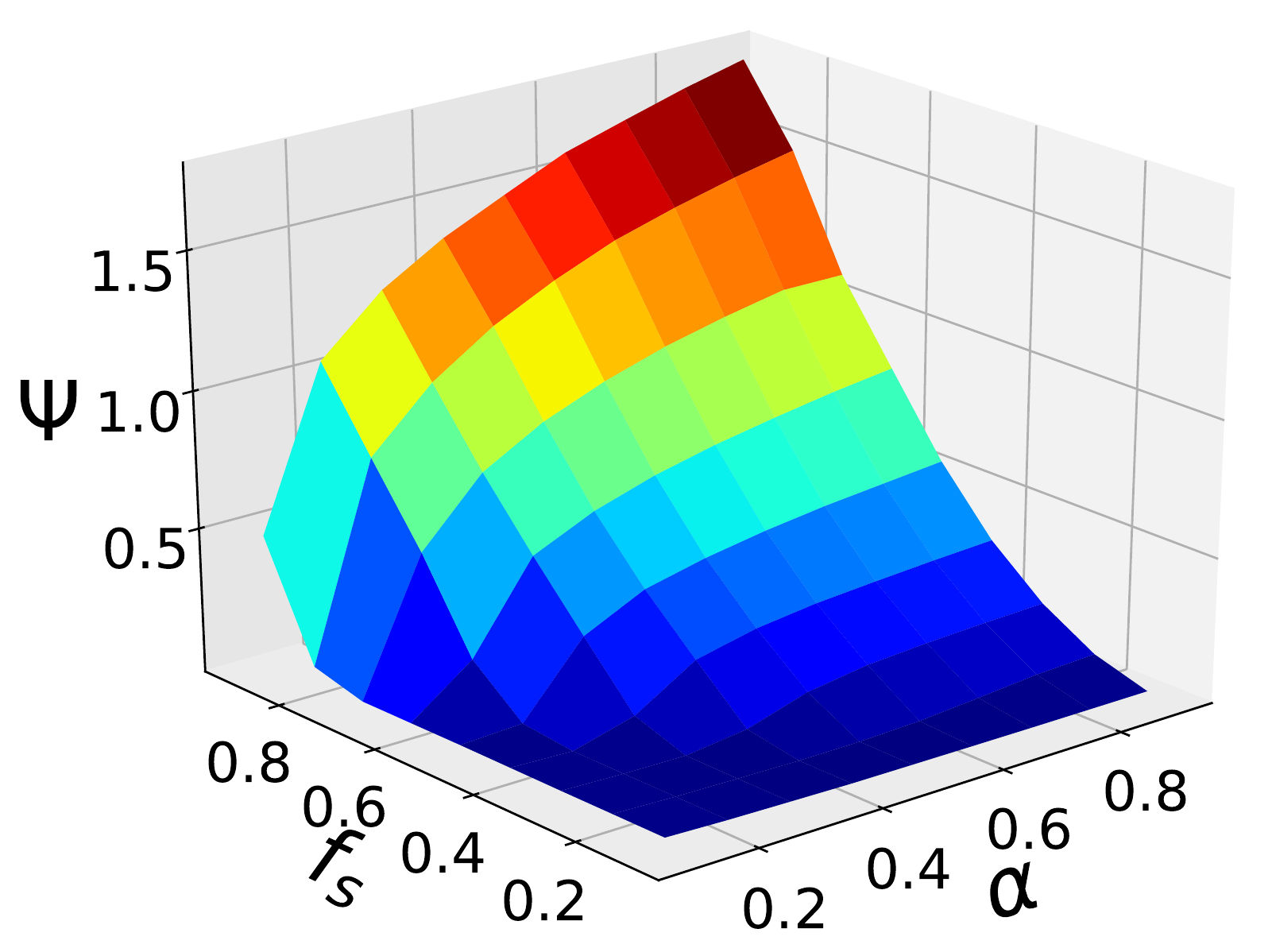, width=0.5\linewidth} &
\epsfig{figure=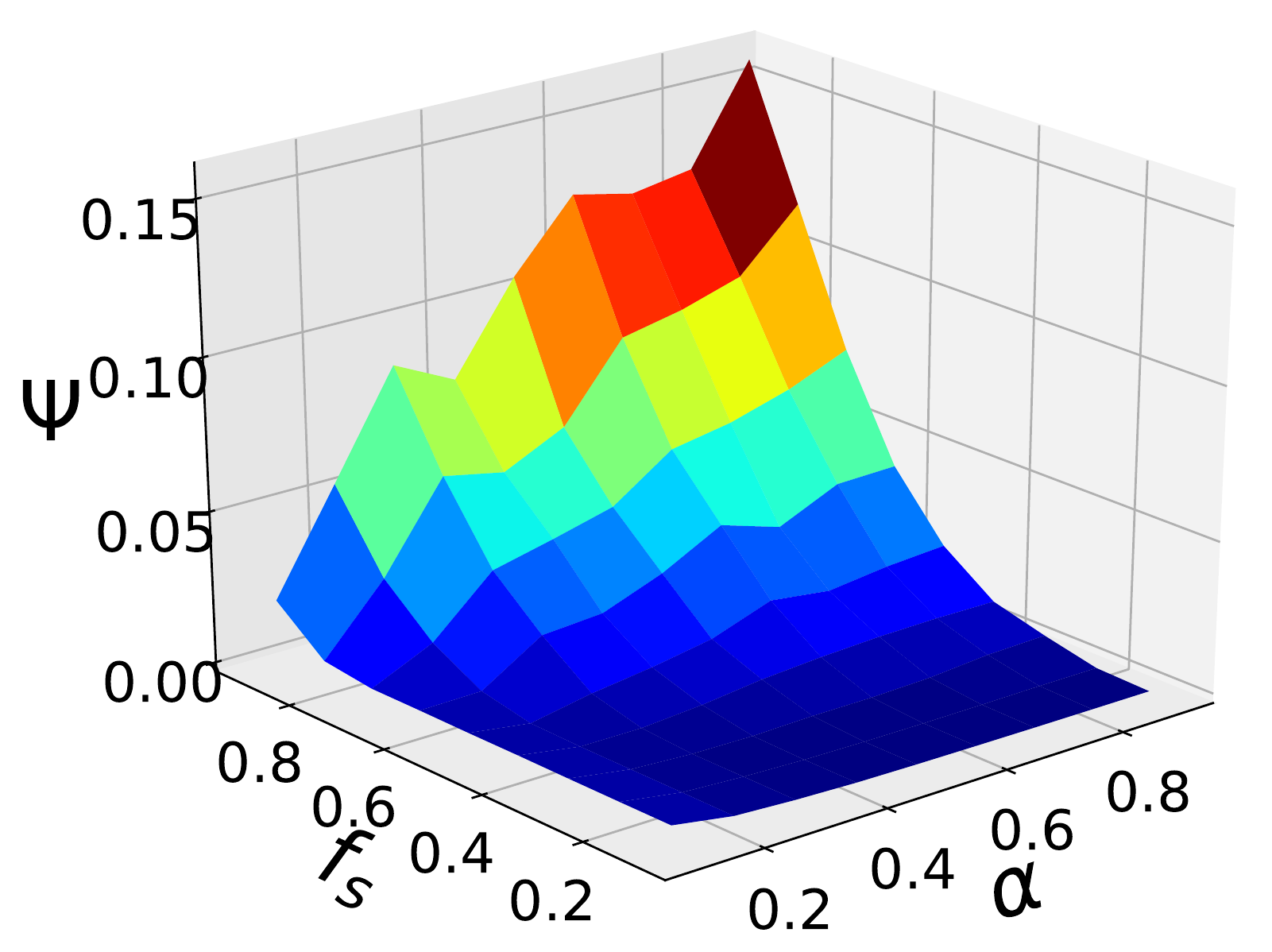, width=0.5\linewidth} \\
$(\gamma_r,\gamma)=(1,2)$ & $(\gamma_r,\gamma)=(2,2)$ 
\end{tabular}
}
\caption{
The fractional difference $\psi$ between the cost after multiple rounds of selfish re-routing and the optimal cost, as a function of $\alpha$ and $\fs$ for (a) $(\gamma_r, \gamma)=(1,2)$ and (b) $(\gamma_r, \gamma)=(2,2)$, respectively. The smaller the value of $\psi$, the closer the system's final state to the optimum. Simulation results are obtained on random regular graphs with $N=100$ and $k=3$ over 1000 realizations. 
}
\label{3Dplot}
\end{figure}

The results of sections~\ref{sec_ga12} and~\ref{sec_ga22} show that selfish users may benefit the transportation system after one round of re-routing. We go on to examine the impact on the system after multiple rounds of selfish re-routing via simulations. Both scenarios of $(\gamma_r, \gamma)=(1,2)$ and $(2,2)$ will be examined. At time $t=0$, all users follow their recommended routes, which are either the shortest path to the common destination ($\gamma_r=1$) or the path-configuration which minimizes the social cost ($\gamma_r=2$). Then, at each time $t\ge 1$, a fraction $\fs$ of the users are randomly selected and re-routed to a path that minimizes their own individual cost, based on traffic conditions at $t\!-\!1$. 

We find that for small $\fs$ the path configuration converges quickly after a few iterations of selfish re-routing, similar to the example shown in \fig{nasheg}(a), (b) and (c) with $\fs=0.5$. No vehicle switches their path in subsequent rounds of re-routing since all of them are already in a path with the minimal individual cost; the system is effectively in a Nash equilibrium state. On the other hand, when $\fs$ is large, the path configuration may fluctuate repeatedly or run into a limit cycle in the synchronous update schedule, for instance, in the case of daily commuters departing at the same time. An example is shown in \fig{nasheg}(a), (d) and (e) with $\fs=1$.
In this example, users follow their shortest path at time $t=0$. At $t=1$, most users re-route to a path which is relatively un-occupied at $t=0$, leaving the originally busy routes at $t=0$ less used at $t=1$. At time $t=2$, these users switch back to the less occupied links at $t=1$, and repeatedly switch back and forth between the occupied and un-occupied routes. 

To better understand the convergence and alternating re-routing behavior, we show in \fig{nashtimeseries} time series of the social cost $\cH$ for exemplar instances. In \fig{nashtimeseries}(a) we show a low vehicle density case $\alpha=0.1$ where $\cH$ converges in instances with $\fs = 0.1$ and $0.5$ where $(\gamma_r, \gamma)=(1,2)$, and for all three instances with $\fs=0.1, 0.5$ and $0.9$ in the case with $(\gamma_r, \gamma)=(2,2)$. 
To reveal the convergence across instances, we show in \fig{nashtimeseries}(c) the fraction of instances with no change in cost for the final consecutive 2500 steps in simulation totaling 5000 steps. Almost all instances converge except those with a large fraction of selfish users $\fs=0.9$ in the case of $(\gamma_r, \gamma)=(1,2)$. This is possibly due to the large number of synchronous updates. An example of the time series of $\cH$ in this case is shown in \fig{nashtimeseries}(a), where $\cH$ fluctuates vigorously at high values; this case may share similarity with the example shown in \fig{nasheg}(a), (d) and (e). These results suggest that at low vehicle density $\alpha$, most cases of the system arrive at a Nash equilibrium via selfish re-routing, as expected.

For cases with a high vehicle density $\alpha=0.9$, we show the time series of $\cH$ from several instances in \fig{nashtimeseries}(b). The time series of $\cH$ for the instances with $\fs=0.5$ and $0.9$ and $(\gamma_r, \gamma)=(1,2)$ exhibit fluctuations that become more vigorous compared to the corresponding instances in \fig{nashtimeseries}(a). As we can see in \fig{nashtimeseries}(d), while most instances still converge for cases with $(\gamma_r, \gamma)=(2,2)$, the convergence ratio starts to drop rapidly beyond $\fs=0.1$ for the case with $(\gamma_r, \gamma)=(1,2)$. These results suggest that with a high vehicle density $\alpha$, it is more difficult for the system to converge to a Nash equilibrium via selfish re-routing, especially if users start from the shortest path configuration. It also suggests that initial route coordination (i.e. $\gamma_r=2$) plays a role in facilitating convergence.

To further examine the system state after multiple rounds of selfish re-routing, we measure the time-averaged social cost $\langle\cH\rangle_t =\frac{1}{100}\sum_{t=101}^{200}\cH(\vs_t|\gamma)$ and define a quantity $\psi$ given by
\begin{align}
\psi = \frac{\langle\cH\rangle_t - \cH(\vsssg|\gamma)}{\cH(\vsssg|\gamma)},
\end{align}
to compare the time-averaged cost after multiple rounds of selfish re-routing with the optimal social cost. As shown in \fig{3Dplot}, $\psi$ behaves similarly with respect to $\fs$ and $\alpha$ for both cases of $\gamma_r=1,2$, while the values of $\psi$ are larger for $\gamma_r=1$. In both cases, $\psi$ increases gradually with $\alpha$ and $\fs$ from $\psi\approx 0$, which suggest that the system reaches a Nash equilibrium state close to the optimal state for systems with a small vehicle density and a small fraction of selfish users, regardless of the initial state of the system. On the other hand, largely sub-optimal states are obtained if the fraction of selfish users is large.

\subsection{Selfish routing on the England highway network}
\label{sec_england}

\begin{figure*}
\centerline{
\begin{tabular}{ccc}
 \multicolumn{1}{l}{(a)} &  \multicolumn{1}{l}{(b)}  &  \multicolumn{1}{l}{(c)} \\
\epsfig{figure=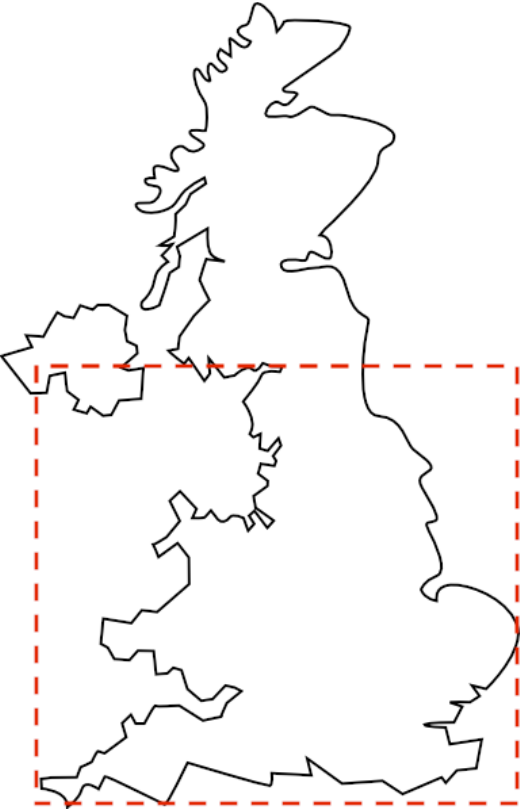, width=0.2\linewidth} &
\epsfig{figure=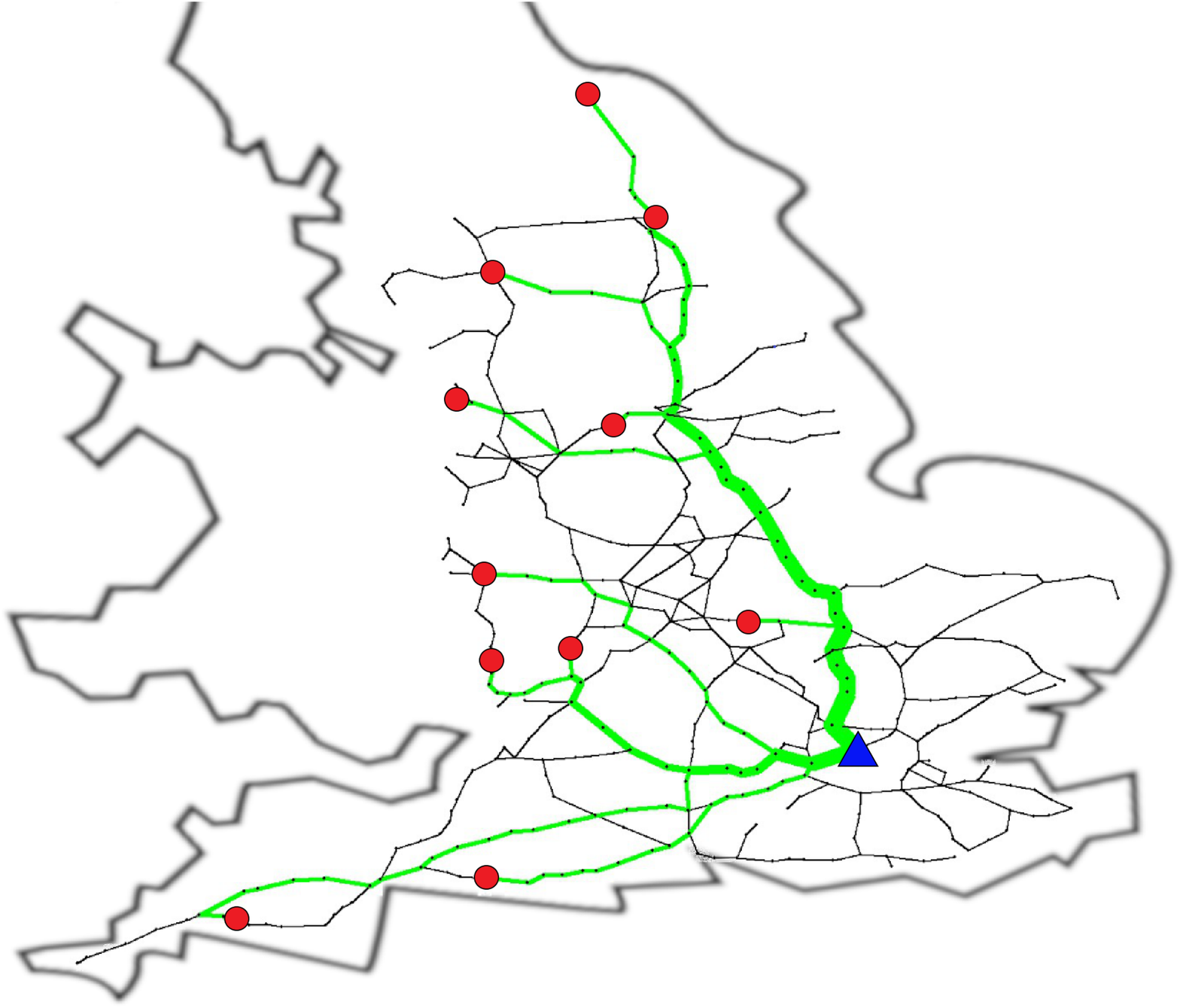, width=0.39\linewidth} &
\epsfig{figure=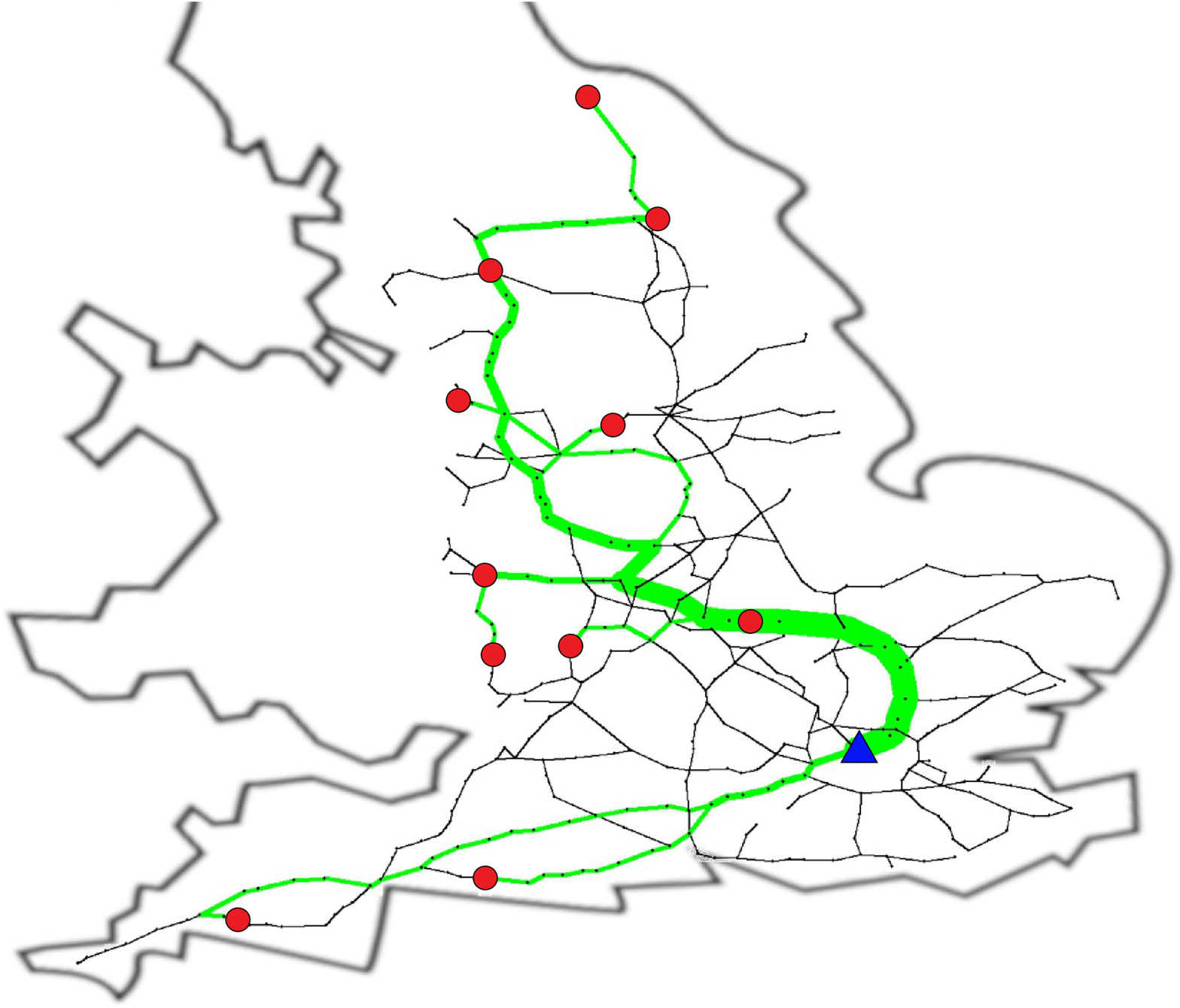, width=0.39\linewidth} 
\\
\end{tabular}
}
\caption{
(a) The region of UK highway network used in our simulation is enclosed by the red rectangle. The network consists of 395 nodes, each of which represents the starting or the ending junction of roads. (b) An example of the configuration of traffic flow resulted from $M=11$ users traveling on the shortest paths from their origin (red filled circles) to the common destination in London (the blue triangle). (c) The configuration of traffic flow after all users re-route. As we can see, all users switch to the initially less occupied route, leaving their original occupied route empty.
}
\label{UK-High-Out}
\end{figure*}

\begin{figure*}
\centerline{
\begin{tabular}{c}
\epsfig{figure=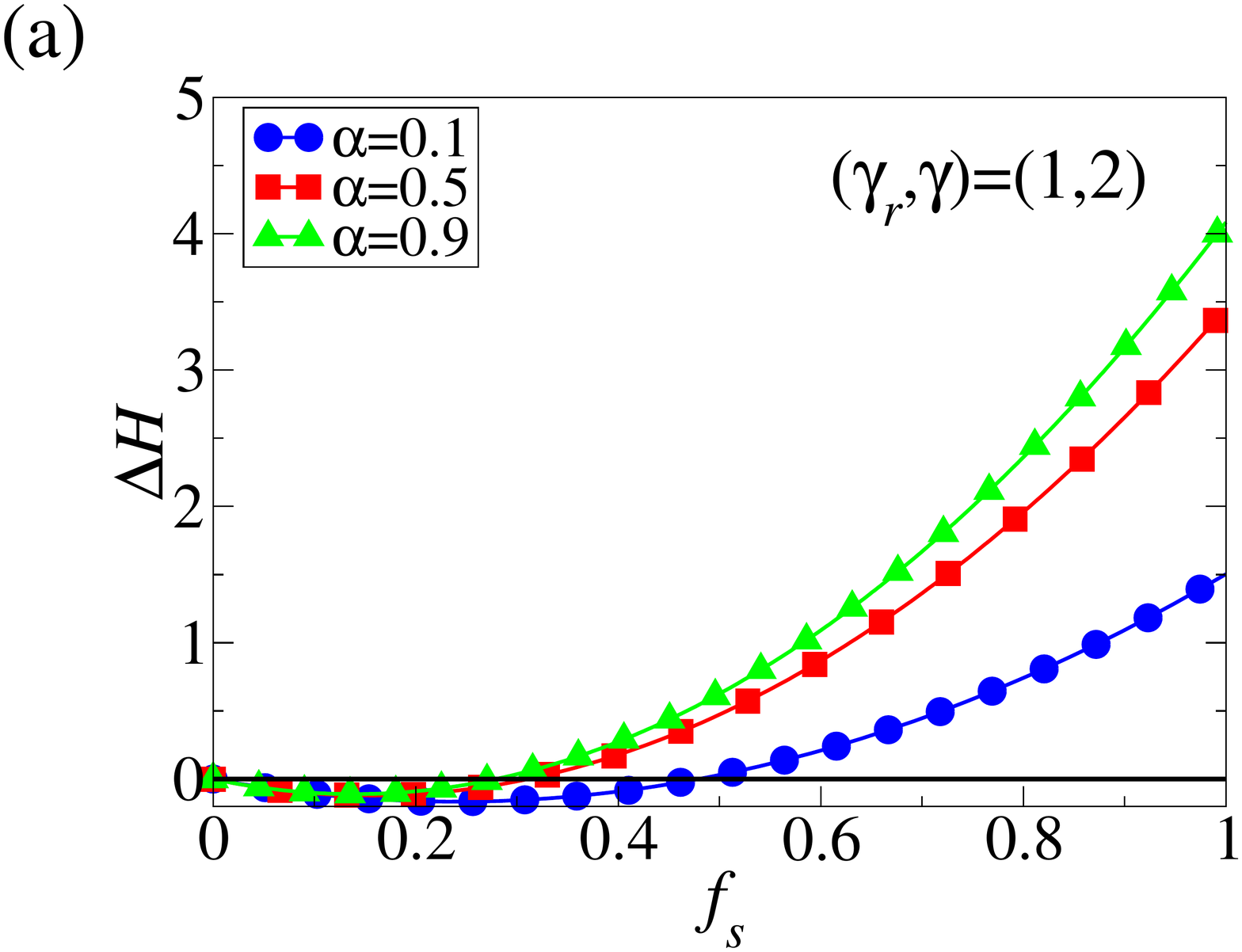, width=0.333\linewidth}
\epsfig{figure=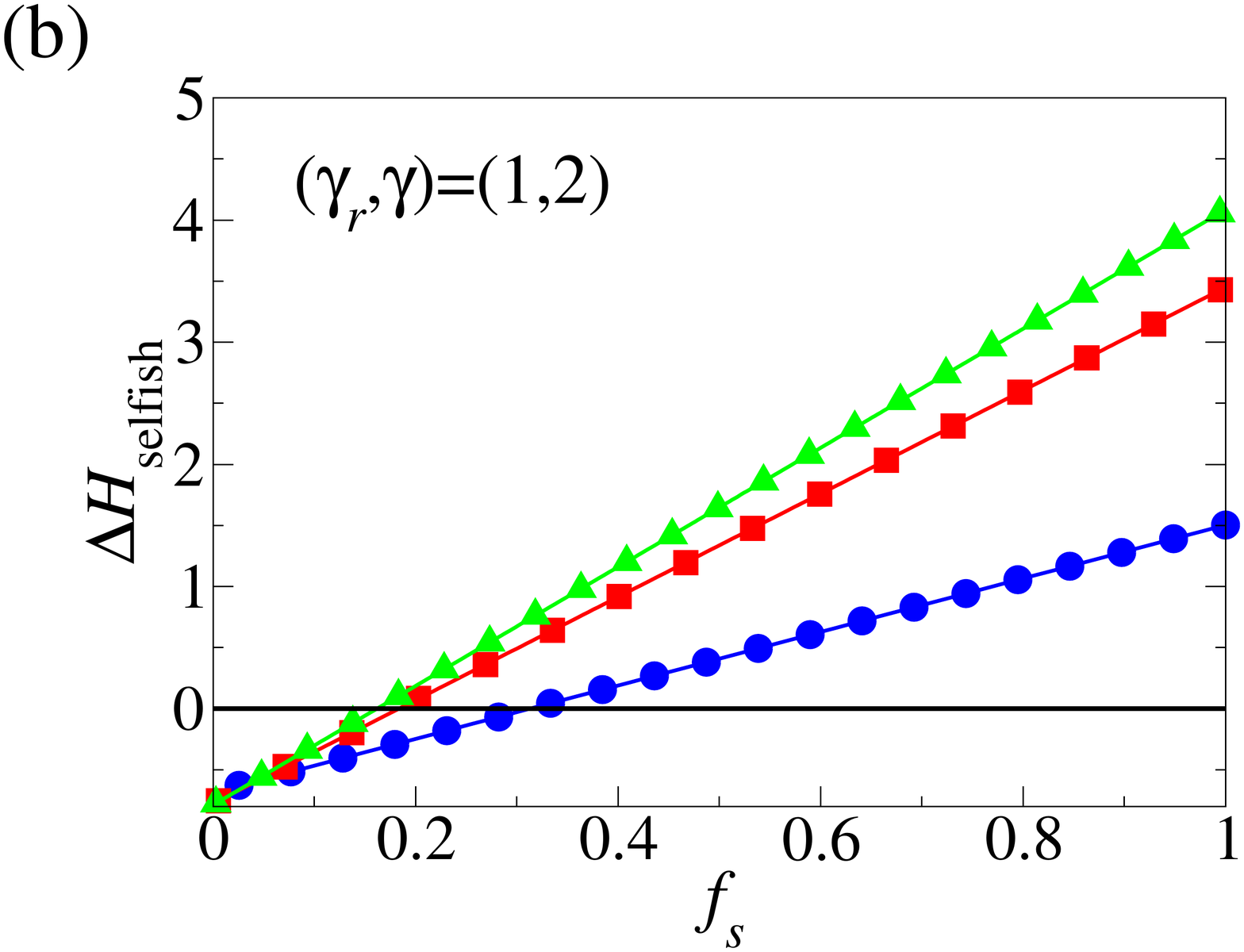, width=0.333\linewidth} 
\epsfig{figure=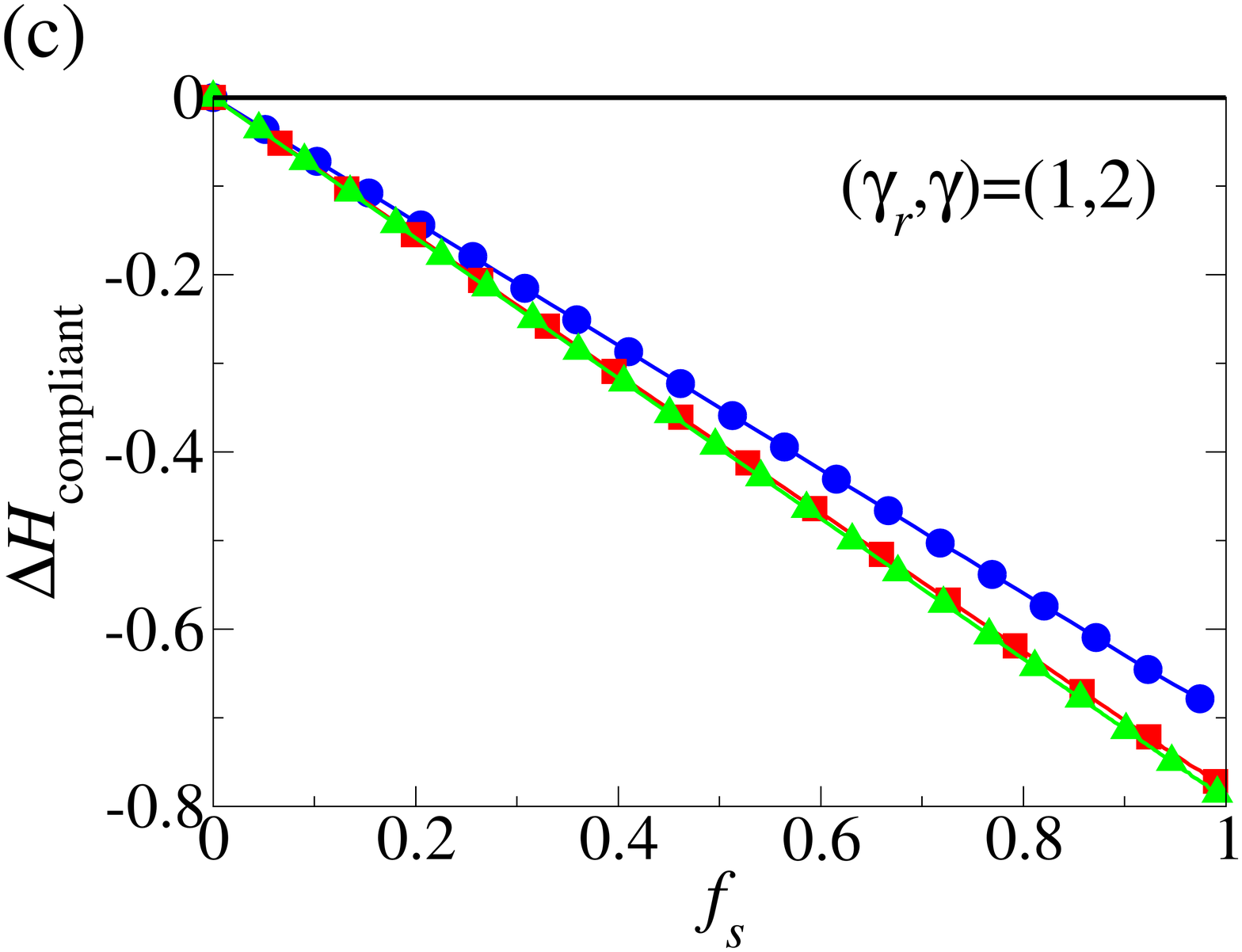, width=0.333\linewidth} \\
\epsfig{figure=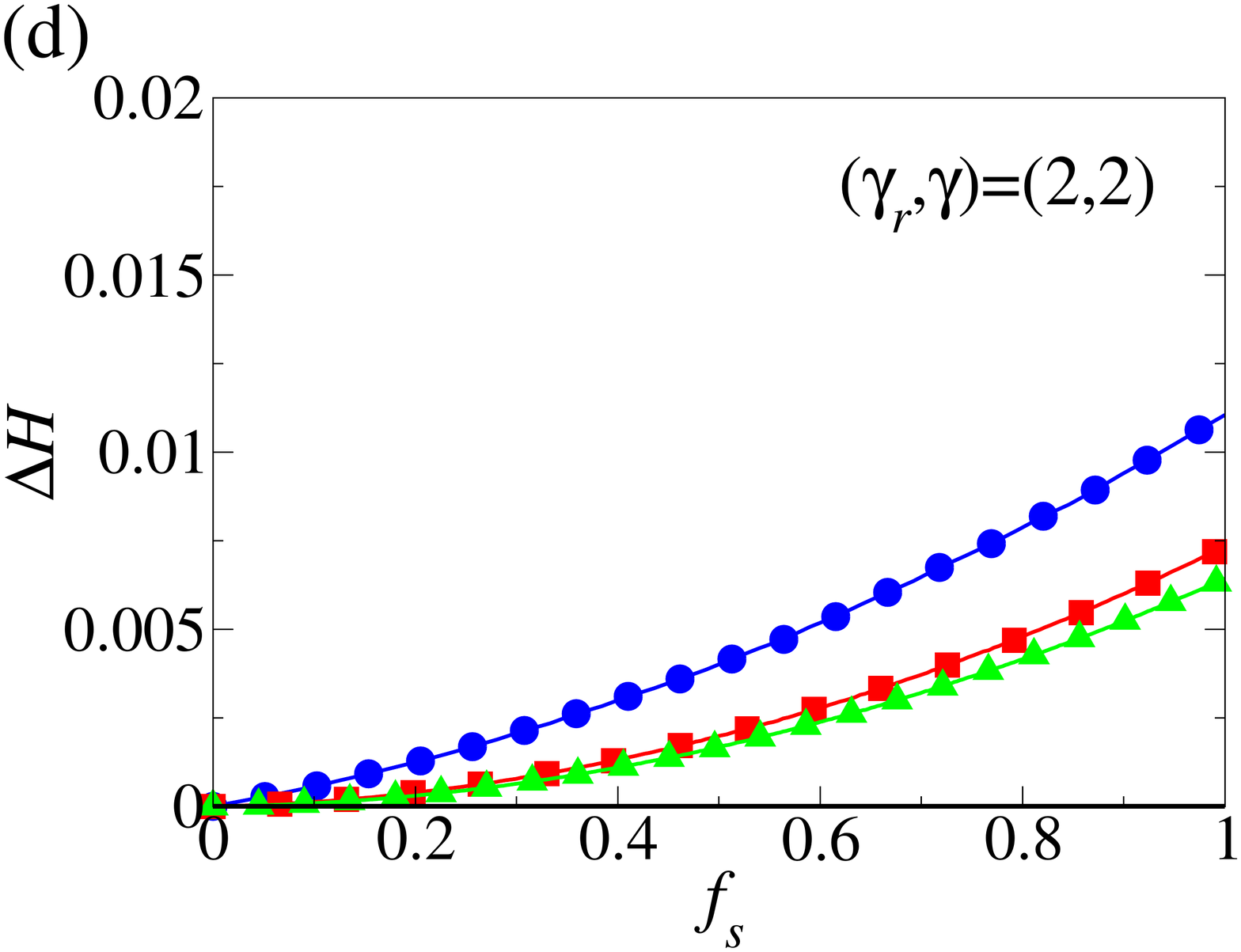, width=0.333\linewidth}
\epsfig{figure=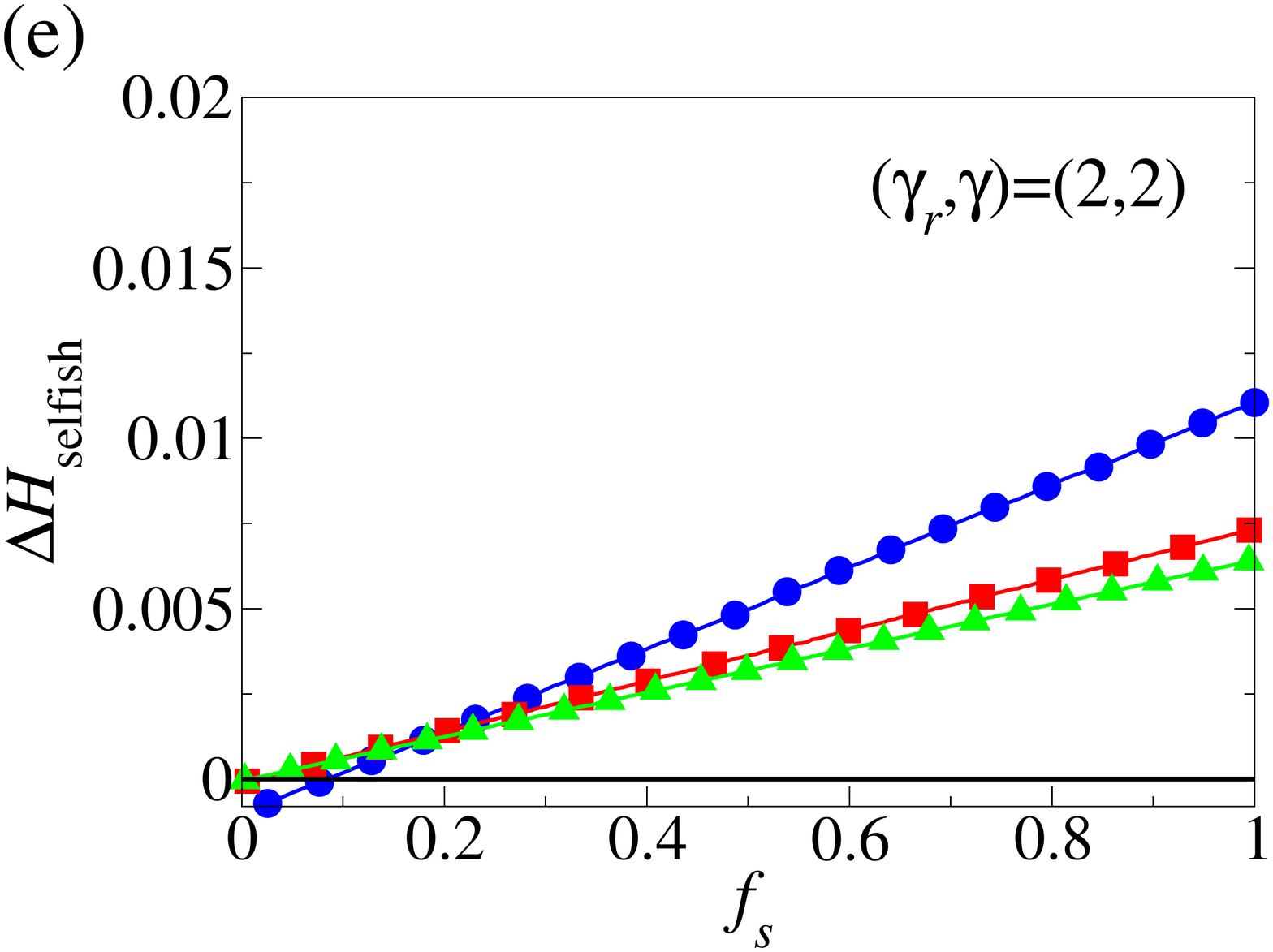, width=0.333\linewidth} 
\epsfig{figure=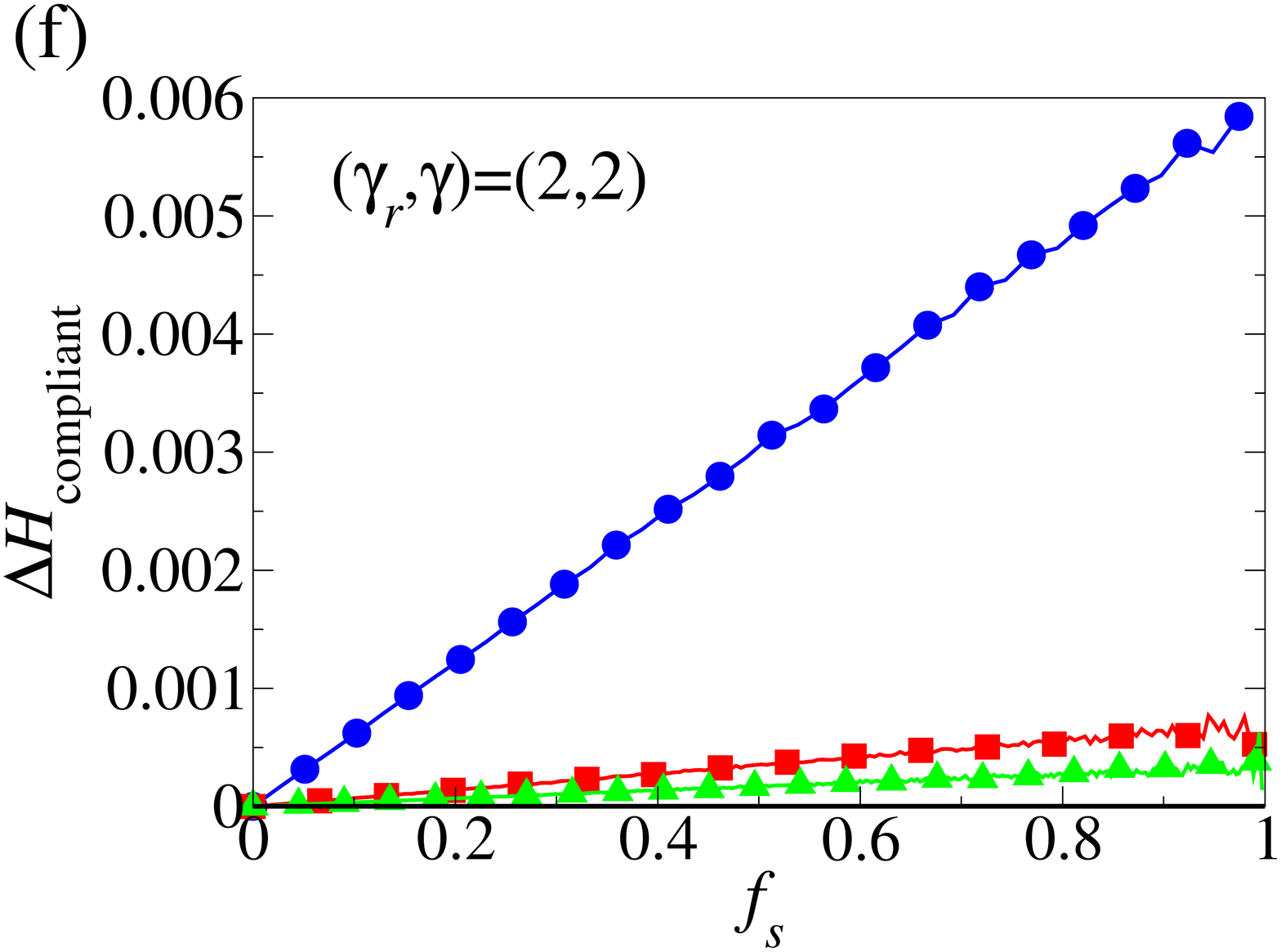, width=0.333\linewidth} \\
\end{tabular}
}
\caption{
The fractional changes (a) $\D$, (b) $\Dself$ and (c) $\Dobed$ of the travel cost averaged over all users, selfish and compliant users, respectively; simulations were carried out for 500 realizations on the England highway network with $(\gamma_r, \gamma)=(1,2)$ and vehicle densities $\alpha=M/N=0.1, 0.5$ and $0.9$. The corresponding simulation results for cases with $(\gamma_r, \gamma)=(2,2)$ are shown in (d), (e) and (f). 
}
\label{UK-FC}
\end{figure*}

\begin{figure}
\centerline{
\begin{tabular}{c}
\epsfig{figure=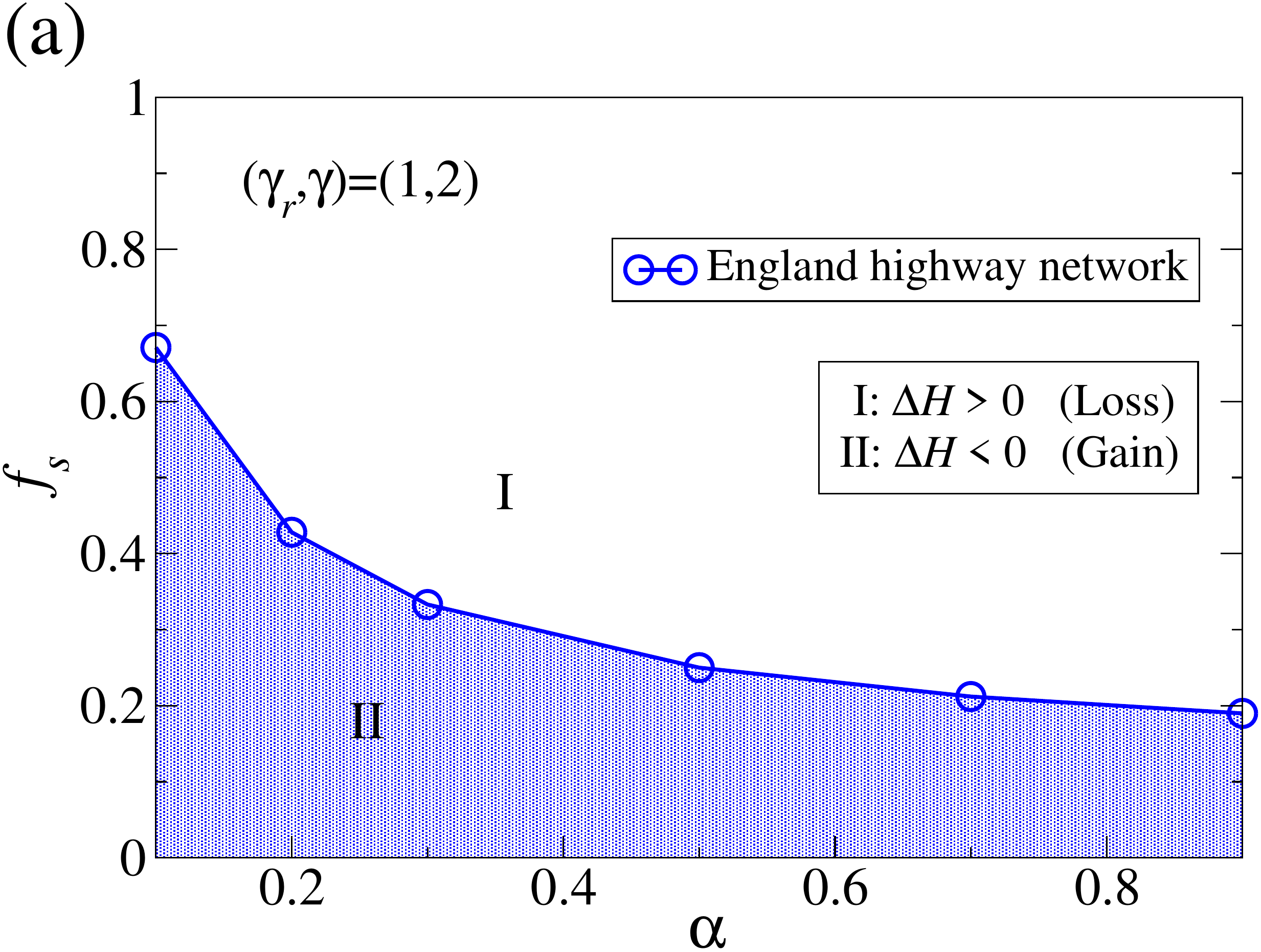, width=0.9\linewidth} \\
\epsfig{figure=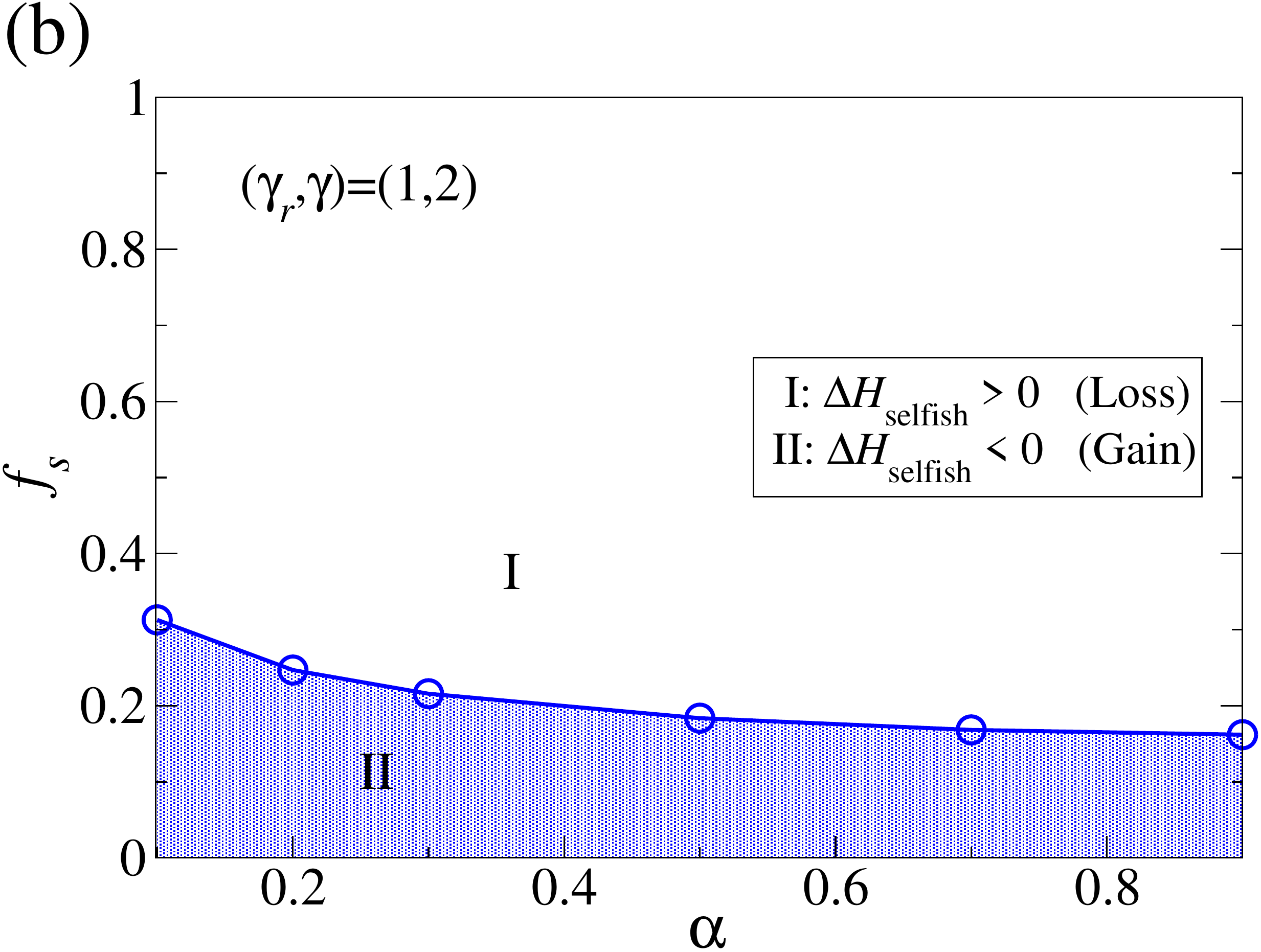, width=0.9\linewidth} \\
\epsfig{figure=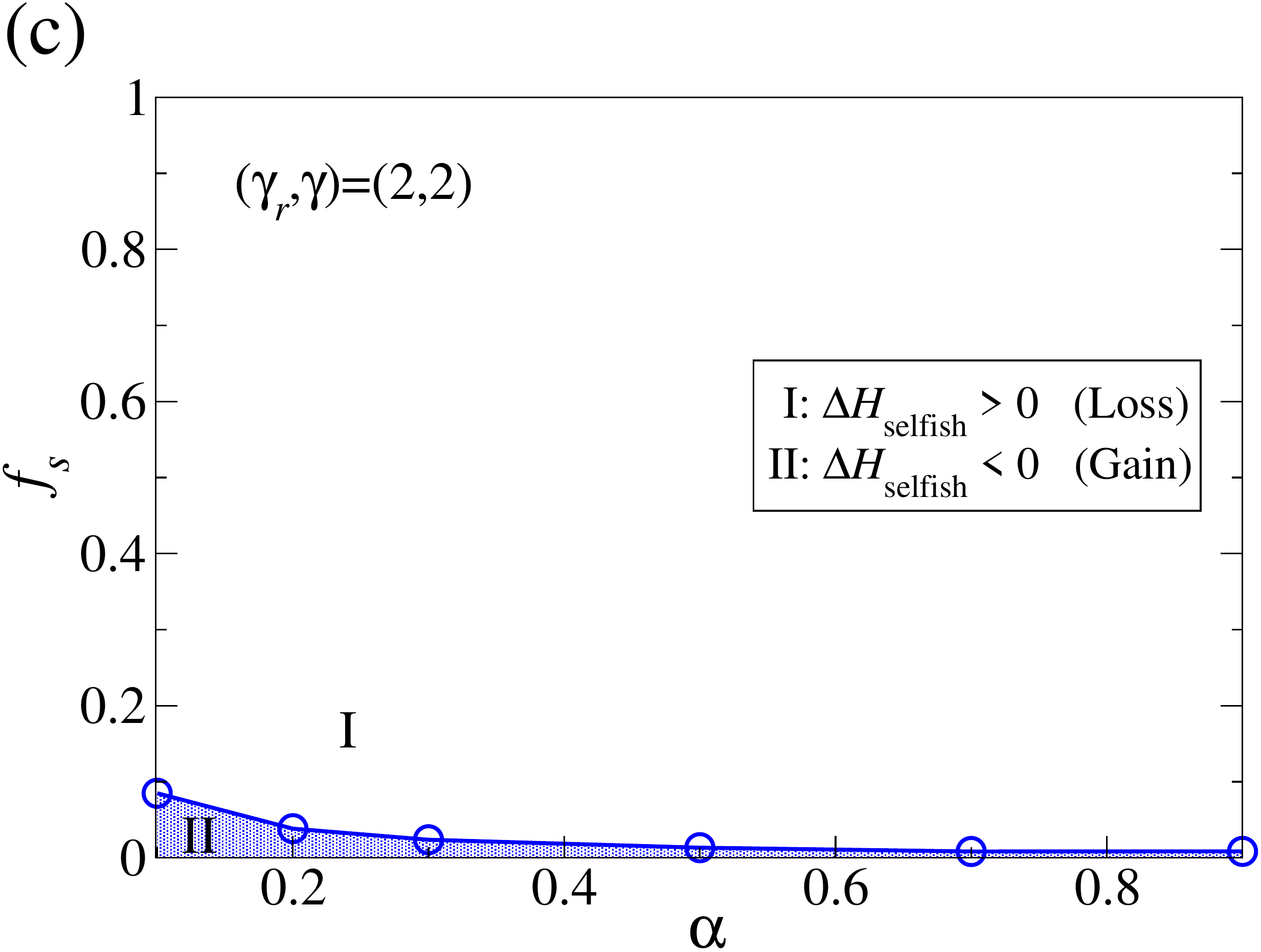, width=0.9\linewidth} \\
\end{tabular}
}
\caption{
The regimes with negative or positive (a) $\D(1,2)$, (b) $\Dself(1,2)$ and (c) $\Dself(2,2)$ in the simulations on the England highway network are shown in terms of the fraction of selfish users $\fs$ and vehicle density $\alpha$.
}
\label{UK-phase}
\end{figure}

\begin{figure}
\centerline{
\begin{tabular}{cc}
 \multicolumn{1}{l}{(a)} &  \multicolumn{1}{l}{(b)}  \\
\epsfig{figure=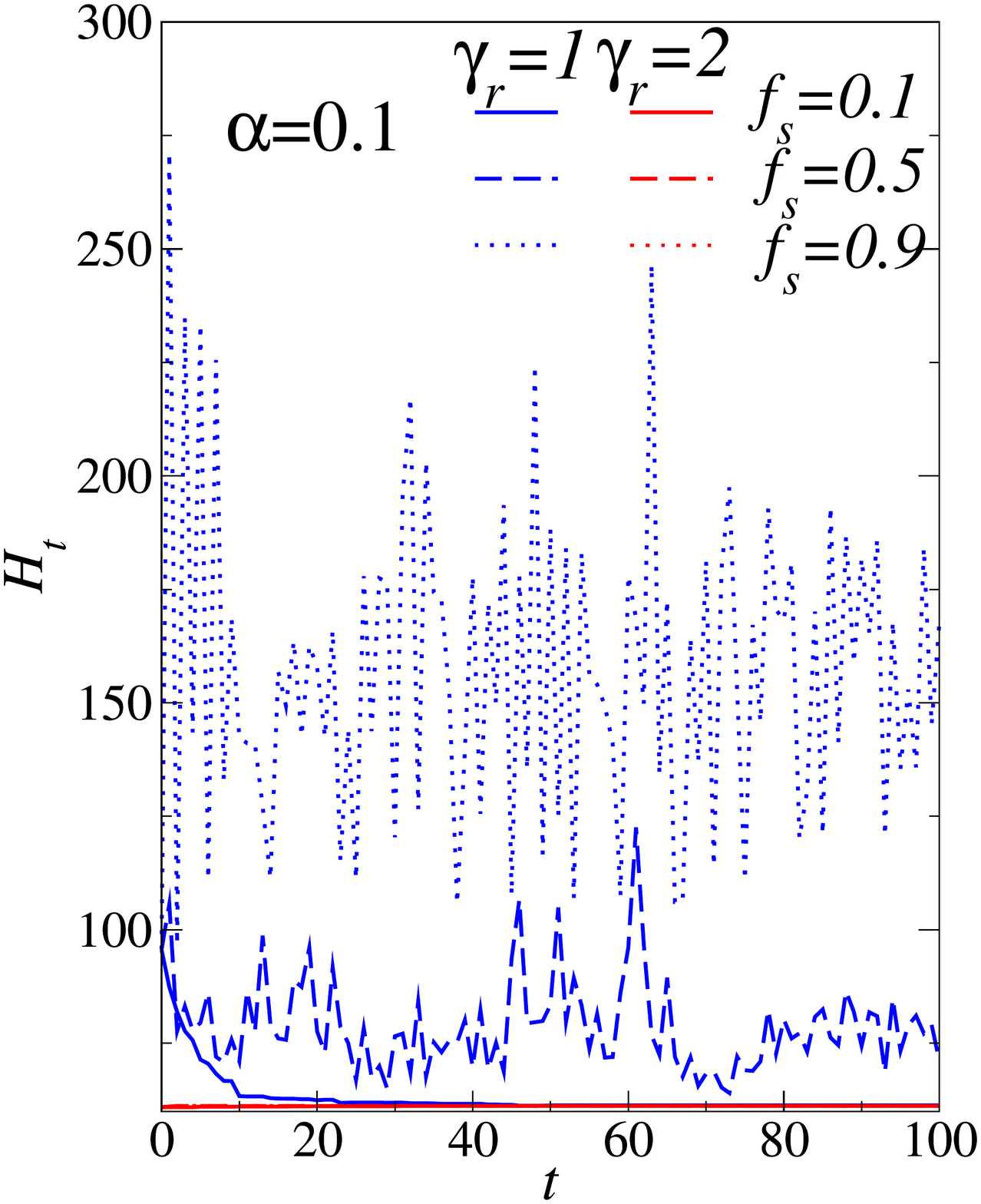, width=0.49\linewidth} &
\epsfig{figure=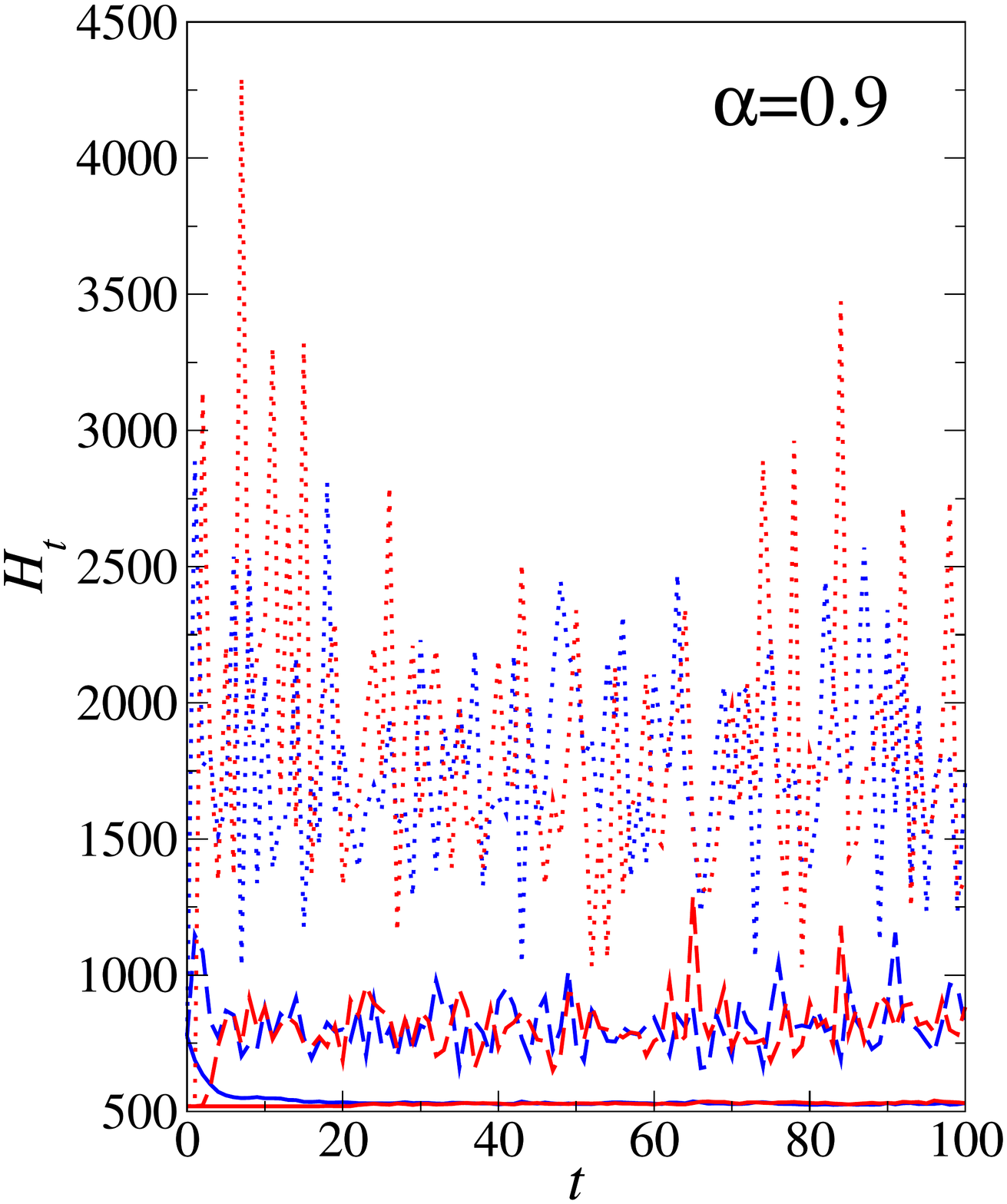, width=0.49\linewidth} \\
\end{tabular}
}
\caption{
The social travel cost $\cH$ on the England highway network as a function of $t$, the rounds of selfish re-routing, measured on specific instances with $\gamma_r =1,2$ and various $\fs$ values for (a) $\alpha=0.1$ and (b) $\alpha=0.9$.
}
\label{UK-nashtimeseries}
\end{figure}

\begin{figure}
\centerline{
\begin{tabular}{cc}
\multicolumn{1}{l}{(a)} &  \multicolumn{1}{l}{(b)} \\
\epsfig{figure=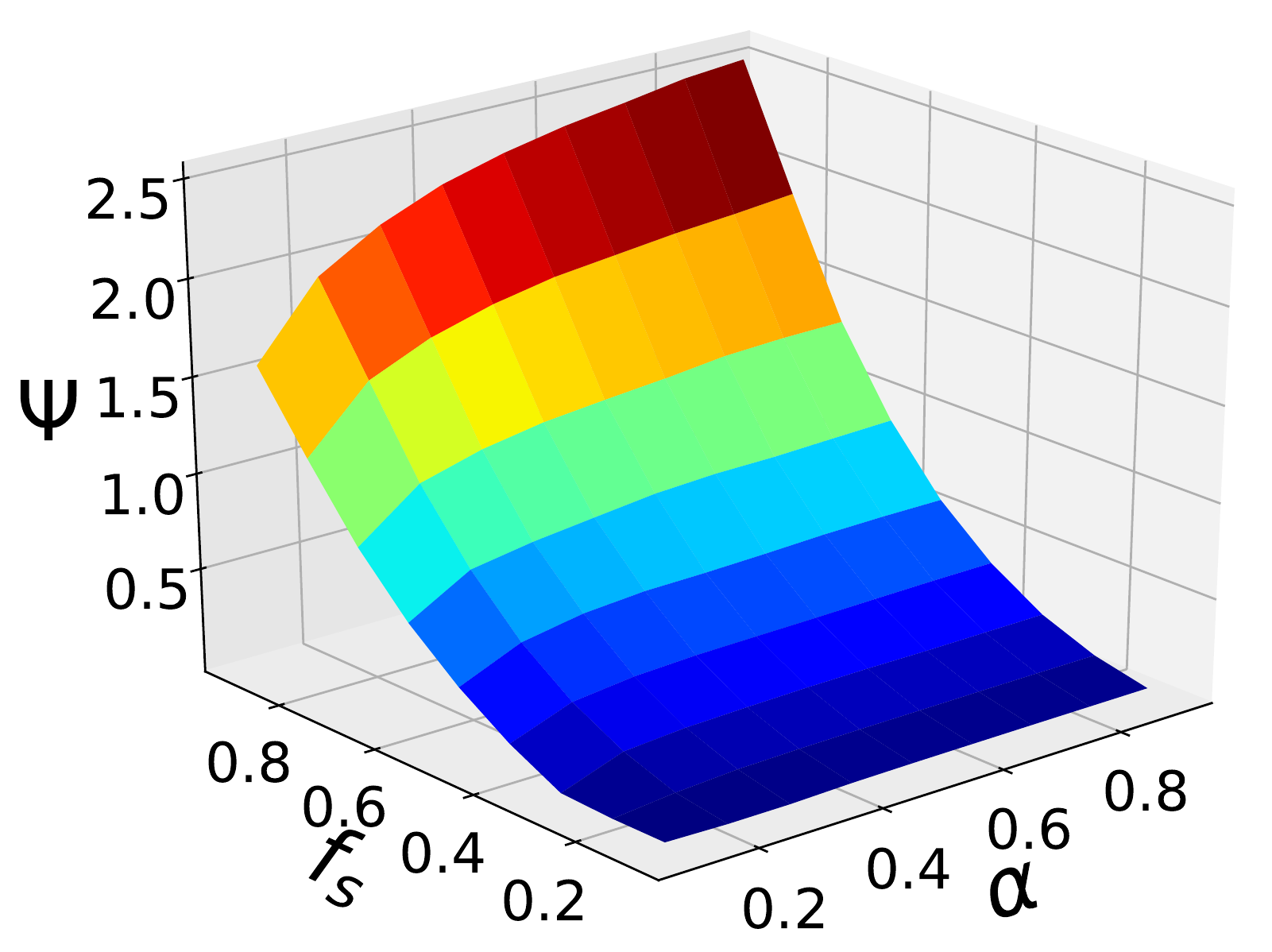, width=0.5\linewidth} &
\epsfig{figure=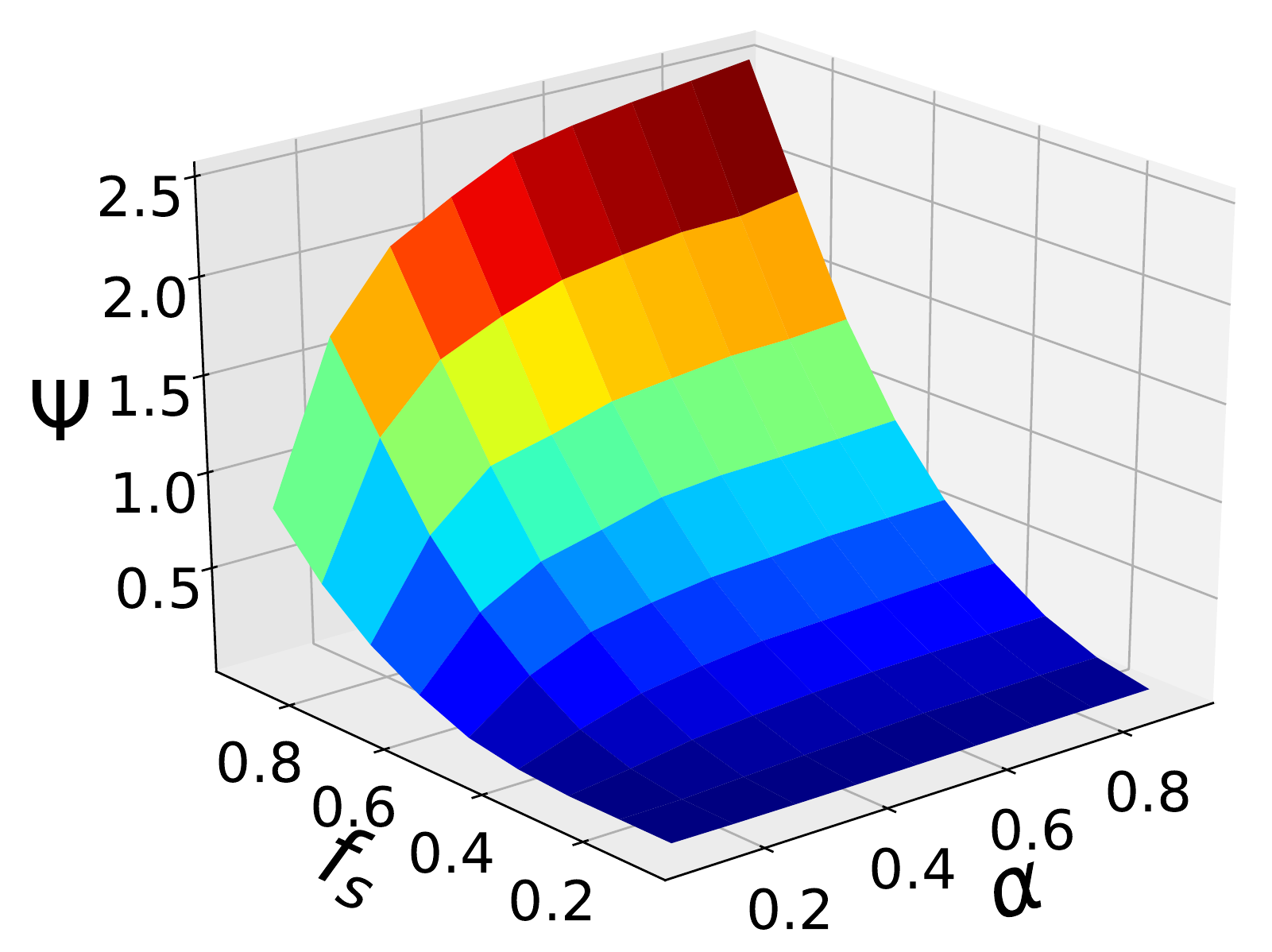, width=0.5\linewidth} \\
$(\gamma_r,\gamma)=(1,2)$ & $(\gamma_r,\gamma)=(2,2)$ 
\end{tabular}
}
\caption{
The fractional difference $\psi$ between the cost after multiple rounds of selfish re-routing and the optimal cost on the England highway network, as a function of $\alpha$ and $\fs$ for (a) $(\gamma_r, \gamma)=(1,2)$ and (b) $(\gamma_r, \gamma)=(2,2)$ respectively. The simulation results are obtained over 500 realizations. 
}
\label{UK-3Dplot}
\end{figure}

To reveal the impact of selfish re-routing on realistic transportation network, we simulate our model on the England highway network with 395 nodes, each representing a starting or ending highway junction based on the data in \cite{UK-Network}. As shown in \fig{UK-High-Out}(a), for simplicity, we create a node in the location of London to serve as the common destination, and allocate an identical weight on all links. An example of shortest path configuration on this highway network with $M=11$ users is shown in \fig{UK-High-Out}(b); the path configuration after all the 11 users re-routed is shown in \fig{UK-High-Out}(c).

In \fig{UK-FC}, we show $\D$, $\Dself$ and $\Dobed$, i.e. the change in cost averaged over all users, selfish users and compliant users, respectively, after one round of re-routing with 500 realizations of simulations. The corresponding parameter regimes when all and  selfish users gain are shown in \fig{UK-phase}. Remarkably, although the topological characteristics of the highway network are different from those in random regular graphs,
the results obtained on the England highway network are similar to those observed in random regular graphs and predicted by the theory, as shown in \fig{FC} and \fig{phase} for both cases of $(\gamma_r, \gamma)=(1,2)$ and $(2,2)$. It also implies that although the analytical results in Sec.~\ref{sec_analytic} are derived based on tree topologies, they capture qualitatively the impact of selfish route decisions on real transportation networks.

We also show simulation results of multiple rounds of selfish re-routing on the highway network in \fig{UK-nashtimeseries} and \fig{UK-3Dplot}. Remarkably, these results are also similar to those observed in random regular graphs in \fig{nashtimeseries} and \fig{3Dplot}, validating the efficacy of our approach. Nevertheless, as we can see in \fig{UK-nashtimeseries}, that the specific topology makes it more difficult for repetitive selfish-routing to converge as more instances show a fluctuating $\cH$ value. In \fig{UK-3Dplot}, we see that $\psi$ increases with $\fs$ and $\alpha$ in both cases of $\gamma_r = 1,2$, suggesting a small vehicle density and a small fraction of selfish users would lead the system to a close-to-optimal Nash equilibria, similar to that observed in random regular graphs. The above results also suggest that the qualitative behaviors of selfish re-routing are robust against network topologies.

\section{Conclusion}
\label{sec_Conclusion}

We studied a model of transportation networks in which optimized routes are recommended to users from their starting point to a common target destination, say a city center. However, having the global routing suggestion some users choose alternative routes to minimize their individual costs based on the traffic experienced. The cavity approach developed in the studies of spin glasses is employed to analyze the impact caused by the selfish re-routing behavior, for all users as well as separately for the groups of compliant and selfish users. 

As shown by both analytical and simulation results, in the case of un-coordinated transportation networks with users following their shortest path, a small fraction of selfish users may reduce the average cost per vehicle globally and hence benefit the system. Their selfish re-routings exploit less-loaded routes, which emerge due to the randomness in both topology and starting points, freeing up overloaded routes. Nevertheless, when the fraction of selfish users increases, the average cost per users increase and the system suffers due to correlation and congestion that appear due to selfish re-routing. Interestingly, the average cost for compliant users which do not alter their routes \emph{always reduces} by the action of selfish users. On the other hand, in the case of optimized transportation networks with quadratic costs, selfish re-routing increases the average cost for all users as well as for compliant users, as expected. Selfish users themselves may gain if their fraction in the user population is small, but will lose out as it grows. 

Using numerical simulations, we show that Nash equilibria states may result after multiple rounds of selfish re-routing, where users no longer re-route once they have all minimized their costs. The social travel cost per vehicle at the Nash equilibria is close to the value at the optimal states when the vehicle density and the fraction of selfish users are small. Similar results are observed on the simulations on the England highway network. 

Our results reveal the impact of selfish routing decisions on the performance of transportation networks, which shed light on benefit brought by route coordination and the importance of influencing drivers' behaviors. The model can accommodate more realistic costs and additional factors to make it more realistic and helpful for transport engineers. Our work also demonstrates how the cavity method can be used to study game-theoretical problems by accommodating the responses of players. This generalized theoretical framework can be readily adapted to study other problems based on iterative alterations by network participants in response to the state of the system.

\section*{Acknowledgements}
This work is supported by the Research Grants Council of the Hong Kong Special Administrative Region, China (Projects No. EdUHK ECS 28300215, GRF 18304316, GRF 18301217 and GRF 18301119), the EdUHK FLASS Dean's Research Fund IRS12 2019 04418, ROP14 2019 04396, and EdUHK RDO Internal Research Grant RG67 2018-2019R R4015. DS acknowledges support from the Leverhulme trust (RPG-2018-092) and the EPSRC Programme Grant TRANSNET (EP/R035342/1). DS would like to thank Alfredo Braunstein and Luca Dall'Asta for helpful discussions at early stages of this work.

\bibliographystyle{prsty}
\bibliography{TrafficGamePRX_ver7}




\end{document}